\documentclass[12pt]{article}
\usepackage[dvips]{graphicx}
\usepackage{epsfig}
\usepackage{amsmath,amssymb,amsbsy}
\usepackage{amsfonts}
\begin{document}
\thispagestyle{empty}
\begin{center}

{\Large\bf{
New developments in the statistical approach of
parton distributions: tests and predictions up to LHC energies}}
\vskip1.4cm
{\bf Claude Bourrely}
\vskip 0.3cm
Aix-Marseille Universit\'e, D\'epartement de Physique, \\
Facult\'e des Sciences site de Luminy, 13288 Marseille, Cedex 09, France\\
\vskip 0.5cm
{\bf Jacques Soffer}
\vskip 0.3cm
Physics Department, Temple University,\\
1925 N, 12th Street, Philadelphia, PA 19122-1801, USA
\vskip 0.5cm
{\bf Abstract}\end{center}

The quantum statistical parton distributions approach proposed more than one
decade ago is revisited by considering a larger set of recent and accurate
Deep
Inelastic Scattering experimental results. It enables us to improve the
description of the data by means of a new determination of the parton
distributions. This global next-to-leading order QCD analysis leads to a good
description of several structure functions, involving unpolarized parton
distributions and helicity distributions, in a broad range of $x$ and $Q^2$
and
in terms of a rather small number of free parameters. There are several
challenging issues and in particular the confirmation of a large positive
gluon
helicity distribution. The predictions of this theoretical approach will be
tested for single-jet production and charge asymmetry in $W^{\pm}$ production
in $\bar p p$ and $p p$ collisions up to LHC energies, using recent data and
also for forthcoming experimental results.

\vskip 0.5cm


\vskip 0.5cm

\newpage
\section{Introduction}
Deep Inelastic Scattering (DIS) of leptons and nucleons is indeed our main
source of information to study the internal nucleon structure in terms of
parton distributions. Several years ago a new set of parton distribution
functions (PDF) was constructed in the
framework of a statistical approach of the nucleon \cite{bbs1}. For quarks
(antiquarks), the building blocks are the helicity dependent distributions
$q^{\pm}(x)$ ($\bar q^{\pm}(x)$). This allows to describe simultaneously the
unpolarized distributions $q(x)= q^{+}(x)+q^{-}(x)$ and the helicity
distributions $\Delta q(x) = q^{+}(x)-q^{-}(x)$ (similarly for antiquarks). At
the initial energy scale $Q_0^2$, these distributions
are given by the sum of two terms, a quasi Fermi-Dirac function and a helicity
independent diffractive
contribution. The flavor asymmetry for the light sea, {\it i.e.} $\bar d (x) >
\bar
u (x)$, observed in the data is built in. This is simply understood in terms
of the Pauli exclusion principle, based on the fact that the proton contains
two up-quarks and only one down-quark. We predict that $\bar d(x) /\bar u(x)$
must remain above one for all $x$ values and this is a real challenge for our
approach, in particular in the large $x$ region which is under experimental
investigation at the moment. The flattening out of the ratio $d(x)/u(x)$ in the
high $x$ region, predicted by the statistical approach, is another interesting
challenge worth mentioning. The chiral properties of QCD lead to
strong relations between $q(x)$ and $\bar q (x)$.
For example, it is found that the well established result $\Delta u (x)>0 $\
implies $\Delta
\bar u (x)>0$ and similarly $\Delta d (x)<0$ leads to $\Delta \bar d (x)<0$. 
This earlier prediction was confirmed by recent polarized DIS data and it was
also demonstrated that the magnitude predicted by the statistical approach is
compatible with recent BNL-RHIC data on $W^{\pm}$ production (see Section 5).
In addition we found the
approximate equality of the flavor asymmetries, namely $\bar d(x) - \bar u(x)
\sim \Delta \bar u(x) - \Delta \bar d(x)$. Concerning
 the gluon, the unpolarized distribution $G(x,Q_0^2)$ is given in
terms of a quasi Bose-Einstein function, with only {\it one free parameter},
and for simplicity, we were assuming zero gluon polarization, {\it i.e.}
$\Delta
G(x,Q_0^2)=0$, at the initial energy scale. As we will see below, the new
analysis of a larger set of recent accurate DIS data, has forced us to give
up this assumption. It leads to the confirmation of a large positive gluon
helicity
distribution, giving a significant contribution to the proton spin, a major
point which was emphasized in a recent letter \cite{bs14}. 
In our previous analysis all unpolarized and
helicity light quark distributions were depending upon {\it eight
free parameters}, which were determined in 2002 (see Ref.~\cite{bbs1}), from a
next-to-leading (NLO) fit of a small set of accurate DIS data. Concerning
the strange quarks and antiquarks distributions, the statistical approach was
applied using slightly different expressions, with four additional parameters
\cite{bbs2}. Since the first determination of the free parameters, new tests
against experimental (unpolarized and
polarized) data turned out to be very satisfactory, in particular in hadronic
reactions, as reported in Refs.~\cite{bbs3,bbs4,bbsW}.\\
It is crucial to note that the quantum-statistical approach differs from the
usual global parton fitting methodology for the following reasons:\\
i) It incorporates physical principles to reduce the number of free parameters
which have a physical interpretation\\
ii) It has very specific predictions, so far confirmed by the data\\
iii) It is an attempt to reach a more physical picture on our knowledge of the
nucleon structure, the ultimate goal would be to solve the problem of
confinement\\
iv) Treating simultaneously unpolarized distributions and helicity
distributions, a unique siuation in the literature, has the advantage to give
access to a vast set of experimental data, in particular up to LHC energies\\
The paper is organized as follows. In Section 2, we review the main points of
our approach and we describe our method to determine the free parameters 
of the PDF with the set of experimental data we have used. In Section 3, we
exhibit all the unpolarized and helicity distributions we have obtained. In
Section 4, we show the results obtained for the unpolarized DIS
structure functions $F_2^{p,d}(x,Q^2)$ in a wide
kinematic range, compared with the world data. We also consider $e^{\pm}p$
neutral and charged current reactions,
and $\nu(\bar{\nu})p$ charged current reactions. This will be completed by our
analysis of polarized DIS experiments, like double helicity asymmetries on a
proton and on a neutron target.
In Section 5, we present predictions for cross sections and helicity
asymmetries in hadronic collisions, in particular inclusive
single-jet production and $W$ production in $\bar{p}p$ and $pp$ collisions, up
to LHC energies. We give our final remarks and conclusions  in the last
section.\\

\section{Basic review on the statistical parton distributions}
Let us now recall the main features of the statistical approach for building
up
the PDF, as opposed
to the standard polynomial type
parameterizations of the PDF, based on Regge theory at low $x$ and on counting
rules at large $x$.
The fermion distributions are given by the sum of two terms,
a quasi Fermi-Dirac function and a helicity independent diffractive
contribution:
\begin{equation}
xq^h(x,Q^2_0)=
\frac{A_{q}X^h_{0q}x^{b_q}}{\exp [(x-X^h_{0q})/\bar{x}]+1}+
\frac{\tilde{A}_{q}x^{\tilde{b}_{q}}}{\exp(x/\bar{x})+1}~,
\label{eq1}
\end{equation}
\begin{equation}
x\bar{q}^h(x,Q^2_0)=
\frac{{\bar A_{q}}(X^{-h}_{0q})^{-1}x^{\bar{b}_ q}}{\exp
[(x+X^{-h}_{0q})/\bar{x}]+1}+
\frac{\tilde{A}_{q}x^{\tilde{b}_{q}}}{\exp(x/\bar{x})+1}~,
\label{eq2}
\end{equation}
at the input energy scale $Q_0^2=1 \mbox{GeV}^2$. We note that the diffractive
term is absent in the quark helicity distribution $\Delta q$ and in the quark
valence contribution $q - \bar q$.\\
In Eqs.~(\ref{eq1},\ref{eq2}) the multiplicative factors $X^{h}_{0q}$ and
$(X^{-h}_{0q})^{-1}$ in
the numerators of the non-diffractive parts of the $q$'s and $\bar{q}$'s
distributions, imply a modification
of the quantum statistical form, we were led to propose in order to agree with
experimental data. The presence of these multiplicative factors was justified
in our earlier attempt to generate the transverse momentum dependence (TMD)
\cite{bbs5}, which was revisited recently \cite{bbs6}.
The parameter $\bar{x}$ plays the role of a {\it universal temperature}
and $X^{\pm}_{0q}$ are the two {\it thermodynamical potentials} of the quark
$q$, with helicity $h=\pm$. They represent the fundamental characteristics of
the model. Notice the change of sign of the potentials
and helicity for the antiquarks \footnote{~At variance with statistical
mechanics where the distributions are expressed in terms of the energy, here
one uses
 $x$ which is clearly the natural variable entering in all the sum rules of
the
parton model.}.\\
For a given flavor $q$ the corresponding quark and antiquark distributions
involve the free parameters, $X^{\pm}_{0q}$, $A_q$, $\bar {A}_q$,
$\tilde {A}_q$, $b_q$, $\bar {b}_q$ and $\tilde {b}_q$, whose number is reduced to $\it
seven$ by the valence sum rule, $\int (q(x) - \bar
{q}(x))dx = N_q$, where $N_q = 2, 1, 0 ~~\mbox{for}~~ u, d, s$, respectively.

For the light quarks $q=u,d$,  the total number of free parameters is reduced
to $\it eight$ by taking, as in Ref. \cite{bbs1}, $A_u=A_d$, $\bar {A}_u =
\bar
{A}_d$, $\tilde {A}_u = \tilde {A}_d$, $b_u = b_d$, $\bar {b}_u = \bar {b}_d$
and $\tilde {b}_u = \tilde {b}_d$. For the strange quark and antiquark
distributions, the simple choice made in Ref. \cite{bbs1}
was improved in Ref. \cite{bbs2}, but here they are expressed in terms of $\it
seven$ free parameters.\\
For the gluons we consider the black-body inspired expression
\begin{equation}
xG(x,Q^2_0) = \frac{A_Gx^{b_G}}{\exp(x/\bar{x})-1}~,
\label{eq3}
\end{equation}
a quasi Bose-Einstein function, with $b_G$ being the only free parameter,
since
$A_G$ is determined by the momentum sum rule.
In our earlier works \cite{bbs1,bbs4}, we were assuming that, at the input
energy scale, the polarized gluon,
distribution vanishes, so
\begin{equation}
x\Delta G(x,Q^2_0)=0~.
\label{eq4}
\end{equation}
However as a result of the present analysis of a much larger set of very
accurate unpolarized and polarized DIS data, we must give up this simplifying
assumption. We are now taking

\begin{equation}
 x\Delta G(x,Q^2_0) = \frac {\tilde A_G x^{\tilde b_G}}{(1+ c_G
x^{d_G})}\!\cdot\!\frac{1}{\exp(x/\bar x - 1) } \,.
 \end{equation}
It is clear that we don't have a serious justification of the functional form
of $\Delta G(x,Q^2_0)$. However the above expression shows that it is strongly
related to  $G(x,Q^2_0)$ and therefore constructed by means of a Bose-Einstein
distribution with zero potential. Actually since $\Delta G(x,Q^2_0)=P(x)
G(x,Q^2_0)$ a simpler expression would be $P(x) = Ax^b$, but the additional
term $x^{d_G}$ in the denominator is needed in order to get a reasonable fit
of
the polarized DIS data. To insure that positivity is satisfied we must have
$|P(x)| \leq 1$ (see section 3). However for quarks and antiquarks positivity
is automatically fullfied by construction.\\
To summarize the new determination of all PDF's involves a total of {\it
twenty
one} free parameters: in addition to the temperature $\bar x$ and the exponent
$b_G$ of the gluon distribution, we have {\it eight} free parameters for the
light quarks $(u,d)$, {\it seven} free parameters for the strange quarks and
{\it four} free parameters for the gluon helicity distribution. These
parameters will be determined from a next-to-leading order (NLO) QCD fit of a
large set of accurate DIS data,  unpolarized and polarized structure
functions,
as we will discuss in the following section.

\section{Unpolarized and polarized parton distributions}
In order to determine these parameters we have performed a global NLO QCD
fitting procedure using only DIS data, because it is well known that the
consideration of semi-inclusive DIS data involves uncertainties related to
fragmentation functions. 
For unpolarized DIS we have considered $F_2^{p,d}(x,Q^2)$ from NMC, E665, H1,
ZEUS, neutral and charged current $e^{\pm}p$ cross sections from HERA and
charged current neutrino and anti-neutrino cross sections from CCFR, NuTeV and
CHORUS, which allow to extract $xF_3^{\nu N}(x,Q^2)$. We present in Table 1
the
details of the number of points and corresponding $\chi^2$ for each
experiment,
with a total of 1773 data points for a total $\chi^2$ of 2288.\\
 For polarized DIS we have considered $g_1^{p,d,n}(x,Q^2)$ from HERMES, E155,
SMC, EMC, E143, E154, JLab and COMPASS. We present in Table 2 the details of
the number of points and corresponding $\chi^2$ for each experiment, with a
total of 269 data points for a total $\chi^2$ of 319. \\The PDF QCD evolution
was done in the $\overline {\bar {\mbox{MS}}}$ scheme  using the HOPPET program \cite{hoppet}, the minimization of the
$\chi^2$ was performed with the
CERN MINUIT program \cite{minuit}. For unpolarized and polarized data we work in the General Mass Variable
Flavour Number Scheme (GM-VFNS) \cite{MSTW,thorne} and for the heavy quark we have taken $m_c$=1.275
GeV. For the strong running coupling $\alpha_s(Q^2)$ we took
$\alpha_s(Q^2_0)=0.32$ and we find $\alpha_s(M_z^2)=0.119 \pm 0.001$.
In our calculations $\alpha_s(M_Z^2)$ is not a free parameter, it comes
from the the evolution equations with an initial value $\alpha_s(Q_0^2)=0.32$.
For polarized data we have not introduced additional constraints coming from
the hyperon decay
constants, but we predict at $Q^2 = 2 \mbox{GeV}^2$, F + D = 1.23 $\pm$ 0.03
and 3F - D = 0.57 $\pm$ 0.02, to be compared respectively with the 
experimental values $1.269 \pm 0.003$ and $0.586 \pm 0.031$
\cite{PDG}\cite{airapet2007}.  
Concerning nuclear corrections for data
involving nuclear targets, in the case  of polarized data for $g_1^d$, we have
taken into account the correction due to D-wave. For unpolarized data, there is
an effect for $F_2$ only in the high $x$ region \cite{ABM} that we have not
considered. However for neutrino DIS data on iron target, the ratio of proton
over neutron is included. For the kinematic cuts we have used the values given
by the experiments and we have restricted the data to $Q^2 > 1\mbox{GeV}^2$ and
$x > 10^{-4}$. We are aware that target mass corrections can be included to
improve the data description \cite{ABM}, but in this simple approach they were
not considered. Also we have taken only leading-twist effects. The error bands
were calculated using the standard Hessian matrix method, following the
prescription described in Ref. \cite{MSTW} and we have used the standard choice
of tolerance $\Delta \chi^2$ =1.
\begin{table}[hbp]
\begin{center}
\begin{tabular}{ c c c c c}
\hline
process &$\chi^2$ &$N_{data}$ & $\chi^2$/d.o.f. & $\chi^2$ 2002 \\
\hline\raisebox{0pt}[12pt][6pt]
{$d \sigma$ ($\nu~p$) CCFR \cite{ccfr01}}       & 271    & 172 & 1.57 & *410
\\[4pt]
{$d \sigma$  ($\nu~p$) NuTeV \cite{nutev06}}    & 206    & 177  &1.16 & *390
\\[4pt]
{$d \sigma$  ($\nu~p$) CHORUS \cite{chorus06}}  & 78     & 64  &1.22  & *176
\\[4pt]
{$d \sigma$ ($\bar\nu~p$) CCFR  \cite{ccfr01}}  & 191    & 163 & 1.17 & *318
\\[4pt]
{$d \sigma$ ($\bar\nu~p$) NuTeV \cite{nutev06} }& 153    & 125  & 1.22& *266
\\[4pt]
{$F_2^p$ E665 \cite{e66596}}                    & 24     & 11  & 2.18&  17
\\[4pt]
{$F_2^p$ ZEUS \cite{zeus01}}                    & 26     & 17  &1.53 &  26
\\[4pt]
{$F_2^p$ H1 \cite{h109,h111}}                   & 105    & 70  &1.5 & *292
\\[4pt]
{$F_2^p$ NMC \cite{nmc97}}                      & 12    & 14   & 0.85 &  13
\\[4pt]
{$F_2^d$ NMC \cite{nmc97}}                      & 230    & 155  &1.48 & 274
\\[4pt]
{$F_2^d/F_2^p$ NMC \cite{nmc97b}}               & 259    & 205   & 1.48 & 198
\\[4pt]
{$F_2^p -F_2^n$ NMC \cite{nmc91,nmc94}}         & 17     &   9  & 1.88 & 49
\\[4pt]
{$xF_3^{\nu N}$ CHORUS \cite{chorus06}}         & 65     & 47  &1.38 & *89
\\[4pt]
{$xF_3^{\nu N}$ NuTeV \cite{nutev06}}       & 68    & 49   & 1.38& 100
\\[4pt]
{Charged current $e^+ p$ HERA \cite{hera10}}     & 33     & 32  &1.03 &  *75
\\[4pt]
{Charged current $e^- p$ HERA\cite{hera10} }     & 18     & 31  & 0.58& *24
\\[4pt]
{Neutral current $e^+ p$ HERA \cite{hera10} }    & 343    & 285 &1.20 & *520
\\[4pt]
{Neutral current $e^- p$ HERA\cite{hera10} }     & 185    & 138  & 1.34& *383
\\[4pt]
{$F_L$ H1 \cite{h110}}                           & 5      & 9   & 0.56&  *37
\\[4pt]
\hline\raisebox{0pt}[12pt][6pt]
{Total }                                 & 2288  & 1773  \\[4pt]
\hline
\end{tabular}
\caption {Detailed $\chi^2$ for the cross sections and the unpolarized
structure functions.
In the column $\chi^2$ 2002 the values marked with an asterisk were not fitted
in 2002.}
\label{table1}
\end{center}
\end{table}

%
\clearpage
\newpage
\begin{table}[ht]
\begin{center}
\begin{tabular}{ c c c c c}
\hline
process &$\chi^2$ &$N_{data}$ & $\chi^2$/d.o.f. & $\chi^2$ 2002  \\
\hline\raisebox{0pt}[12pt][6pt]
{$g_1^p$ HERMES \cite{herm05}}        & 34  & 34 & 1  & *36  \\[4pt]
{$g_1^p$ E155 \cite{e15500a}}         & 6   & 8  & 0.75 & 8   \\[4pt]
{$g_1^p$ SMC  \cite{smc98}}           & 25  & 12  & 2.08  &  35  \\[4pt]
{$g_1^p$ EMC \cite{emc88a,emc88b}}    & 9   & 10  & 0.9 & 8   \\[4pt]
{$g_1^p$ E143 \cite{e14398a}}         & 33  & 28  & 1.18 &  34  \\[4pt]
{$g_1^p$ COMPASS \cite{compass10}}    & 16  & 9   & 1.77 & *23  \\[4pt]
{$g_1^n$ SMC \cite{smc98}}            & 4   & 7  & 0.57 & 8  \\[4pt]
{$g_1^n$ E155 \cite{e15500a}}          & 10  & 11  & 0.91 & 12  \\[4pt]
{$g_1^n$ E154 \cite{e15497}}          & 5   & 11   & 0.45 & 6  \\[4pt]
{$g_1^n$ E143  \cite{e14398a}}        & 41  & 27  &1.52  &  43  \\[4pt]
{$g_1^n$ Jlab  \cite{jlab04}}         & 3   & 3  &1  &  *1  \\[4pt]
{$g_1^d$ HERMES \cite{herm05,herm07} }& 43  & 36  & 1.19  &  *70  \\[4pt]
{$g_1^d$ COMPASS \cite{comp10b}}      & 12  & 10  & 1.2 & *30  \\[4pt]
{$g_1^d$ E155 \cite{e15599}}          & 21  & 23  & 0.91  &  *43  \\[4pt]
{$g_1^d$ E143 \cite{e14398a} }        & 34  & 28   & 1.21  & 43  \\[4pt]
{$g_1^d$ SMC \cite{smc98} }           & 23  & 12  & 1.92 &  *37  \\[4pt]
\hline\raisebox{0pt}[12pt][6pt]
{Total } & 319  & 269   \\[4pt]
\hline
\end{tabular}
\caption {Detailed $\chi^2$ for the polarized structure functions
$g_1^{p,d,n}(x,Q^2)$.
In the column $\chi^2$ 2002 the values marked with an asterisk were not fitted
in 2002.}
\label{table2}
\end{center}
\end{table}
The new determination of the PDF \footnote { To compute the unpolarized
distributions and the helicity distributions, a Fortran program is available upon
request.} leads, for the light quarks $(q=u,d)$, to the
following parameters:
\begin{eqnarray}
\nonumber
A_q = 1.943 \pm 0.005,~b_q = 0.471 \pm 0.001~\bar {A}_q = 8.915 \pm 0.050,\\
{}~\bar {b}_q = 1.301 \pm 0.004,~\tilde A_q = 0.147 \pm 0.003, ~\tilde b_q=
0.0431 \pm 0.003
\label{eq9}
\end{eqnarray}
and four potentials 
\begin{eqnarray}
\nonumber
X_{0u}^+= 0.475 \pm 0.001,~X_{0u}^-= 0.307 \pm 0.001,\\ X_{0d}^+= 0.245 \pm
0.001,~ X_{0d}^-= 0.309 \pm 0.001.
\label{eq10}
\end{eqnarray}
Concerning the strange quarks we have  the following parameters:
\begin{eqnarray}
\nonumber
A_s = 28.508\pm 0.005,~b_s = 0.370 \pm 0.002,~\bar {A}_s = 0.0026 \pm
0.0002,\\
\bar {b}_s = 0.201 \pm 0.003,~\tilde A_s = 13.689 \pm 0.050, ~\tilde b_s=
9.065
\pm 0.020,
\label{eq11}
\end{eqnarray}
and two potentials
\begin{equation}
X_{0s}^+= 0.011 \pm 0.001,~X_{0s}^-= 0.015 \pm 0.001.
\label{eq12}
\end{equation}
Finally in the gluon sector, we obtain the following parameters:
\begin{eqnarray}
\nonumber
A_G = 36.778 \pm 0.085,~b_G = 1.020 \pm 0.0014,~\tilde {A}_G = 26.887 \pm
0.050,\\
\tilde {b}_G = 0.163 \pm 0.005,~c_G = 0.006 \pm 0.0005, ~d_G = -6.072 \pm
0.350.
\label{eq8}
\end{eqnarray}
In addition the new universal temperature is $\bar x =0.090 \pm 0.002$.

\begin{figure}[hbp]   
\vspace*{-21.5ex}
\begin{center}
\includegraphics[width=7.5cm]{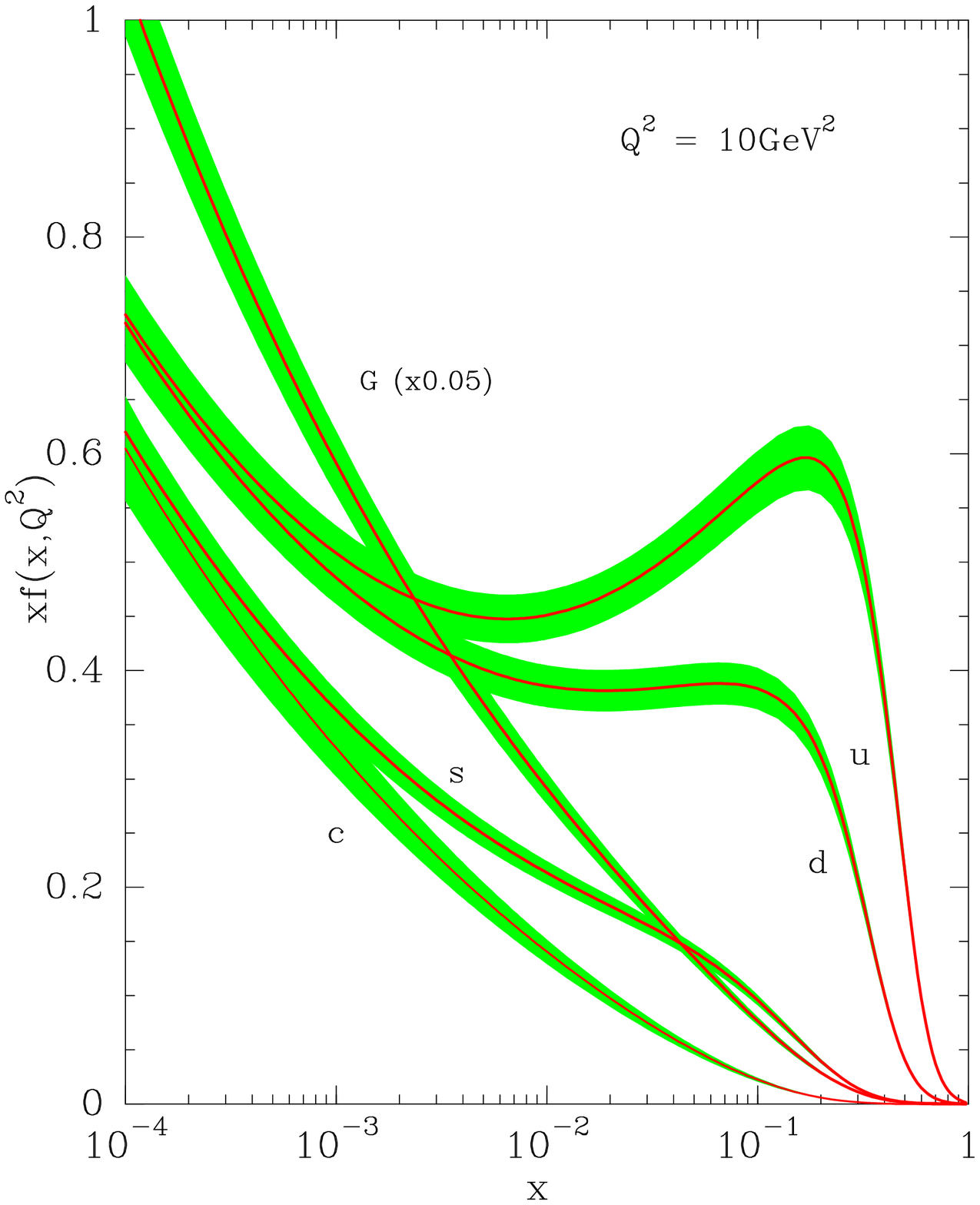}
\vspace*{7.0ex}
\includegraphics[width=7.5cm]{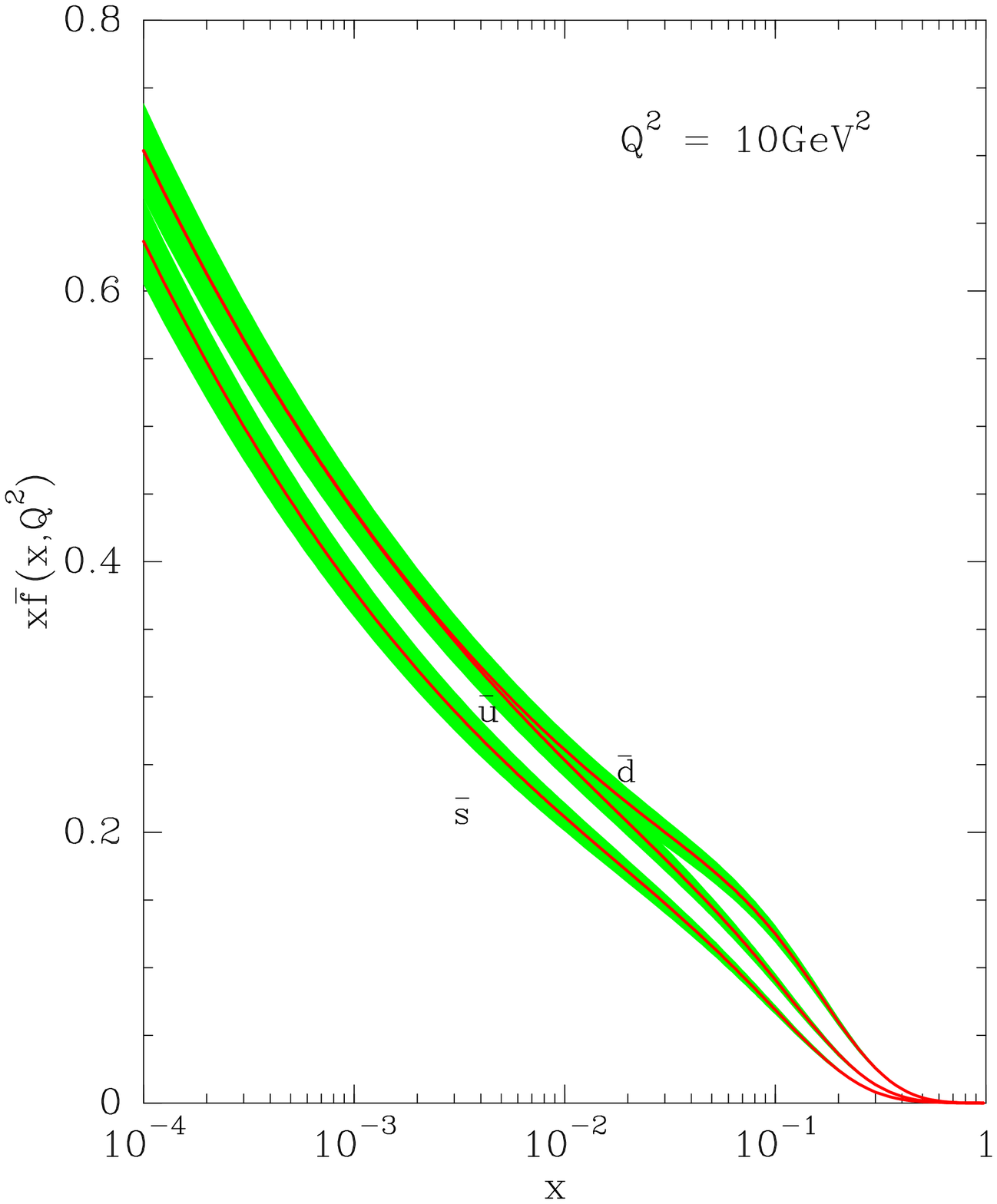}
\vspace*{-5.0ex}
\caption[*]{\baselineskip 1pt
 The different unpolarized parton distributions $xf(x,Q^2)$ ($f=u, d, s, c,
\bar u, \bar d, \bar s ~ \mbox{and} ~  G$)  versus $x$, after NLO QCD
evolution
at $Q^2=10\mbox{GeV}^2$, with the corresponding error bands. The charm
distributions $xc(x,Q^2) = x\bar c(x,Q^2)$ are generated by QCD evolution.}
\label{unpolglobal}
\end{center}
\end{figure}

\begin{figure}[hbp]  
\vspace*{-10.5ex}
\begin{center}
\includegraphics[width=7.5cm]{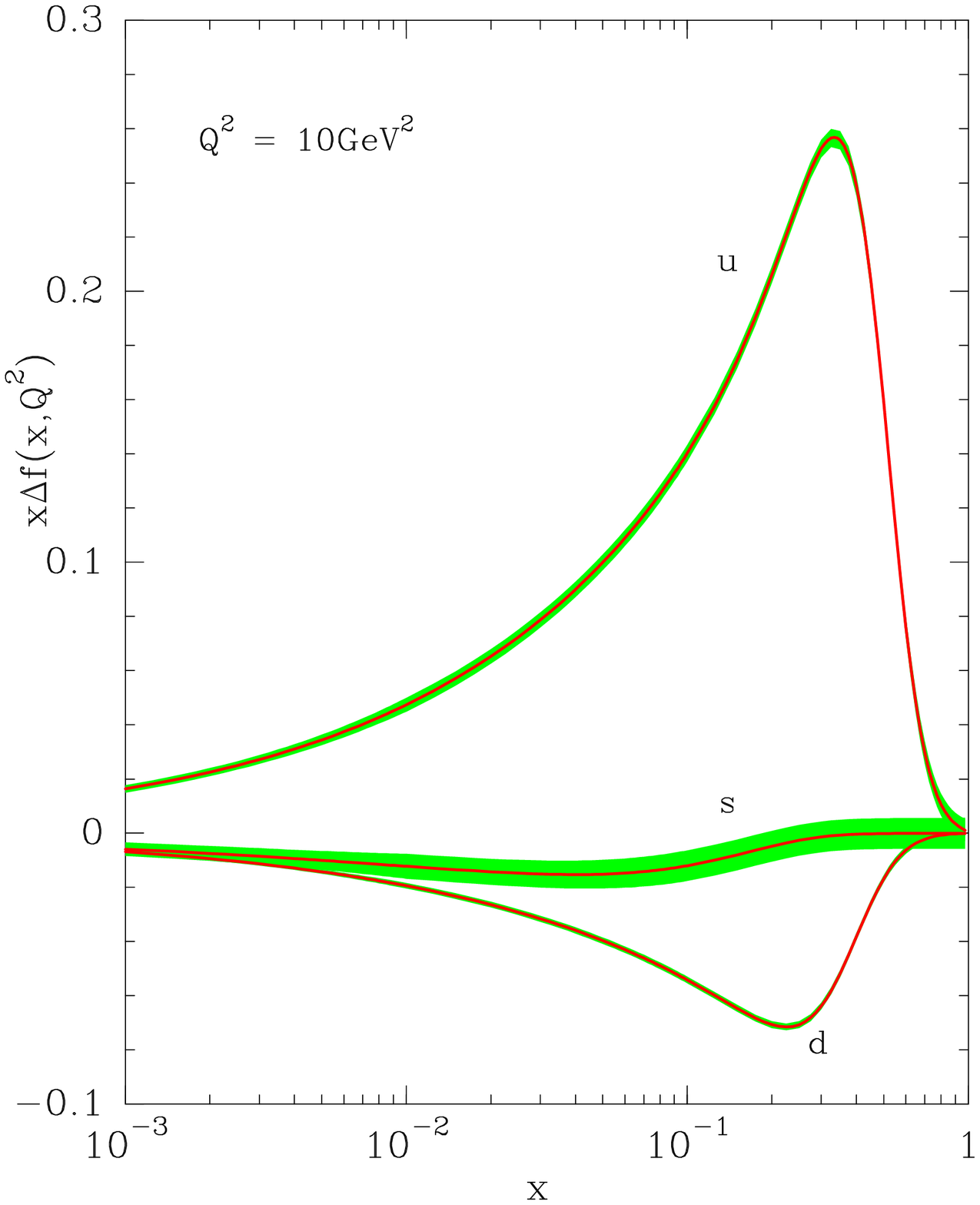}
\includegraphics[width=7.5cm]{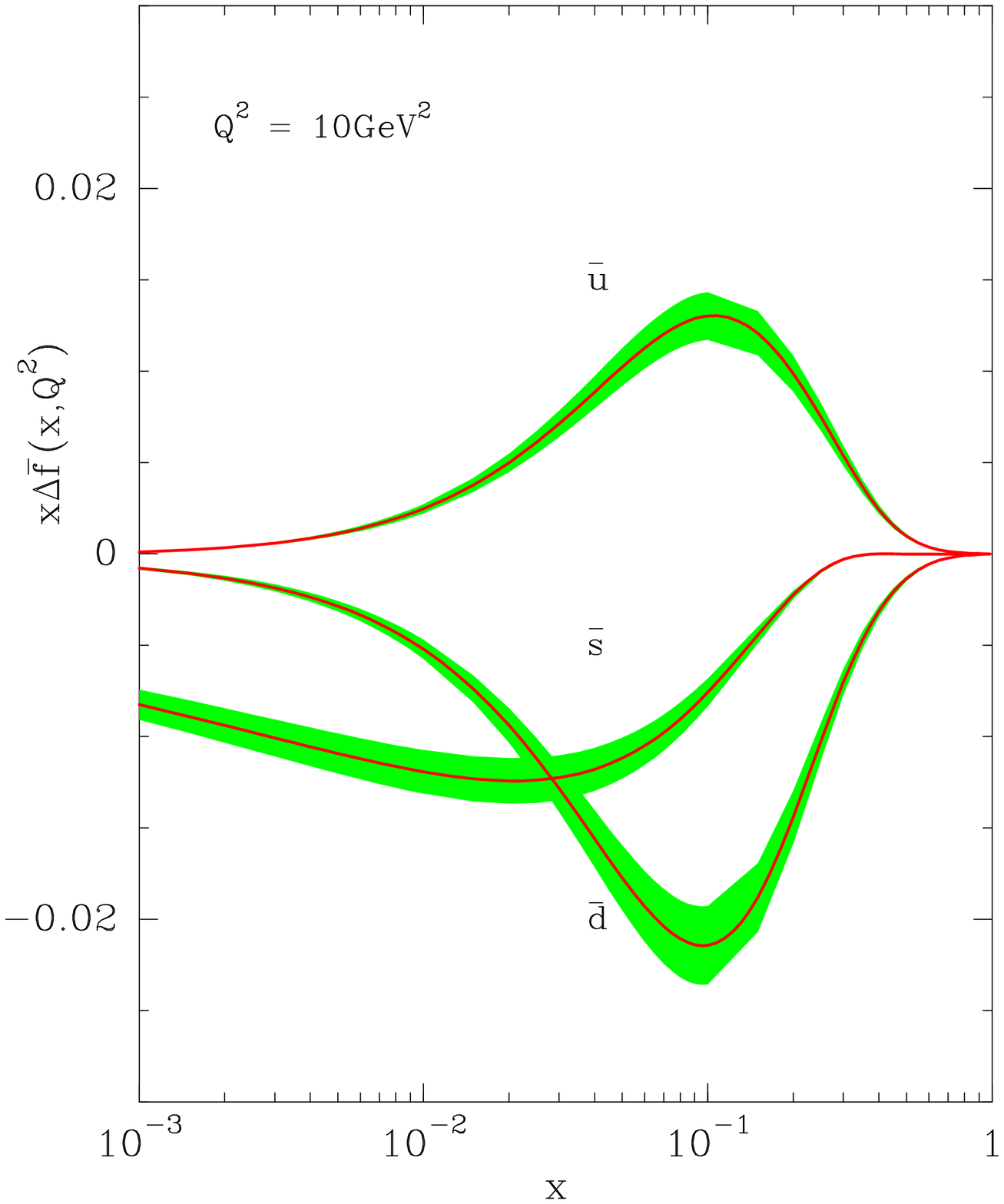}
\caption[*]{\baselineskip 1pt
  The different quark and antiquark helicity distributions $x\Delta f(x,Q^2)$
($f=u, d, s,\bar u, \bar d ~ \mbox{and} ~  \bar s $)  versus $x$, after NLO
QCD
evolution at $Q^2=10\mbox{GeV}^2$, with the corresponding error bands.  The
charm helicity distributions generated by QCD evolution are essentially zero}
\label{polglobal}
\end{center}
\end{figure}

\begin{figure}[hbp]   
\vspace*{-5.5ex}
\begin{center}
 \includegraphics[width=9.5cm]{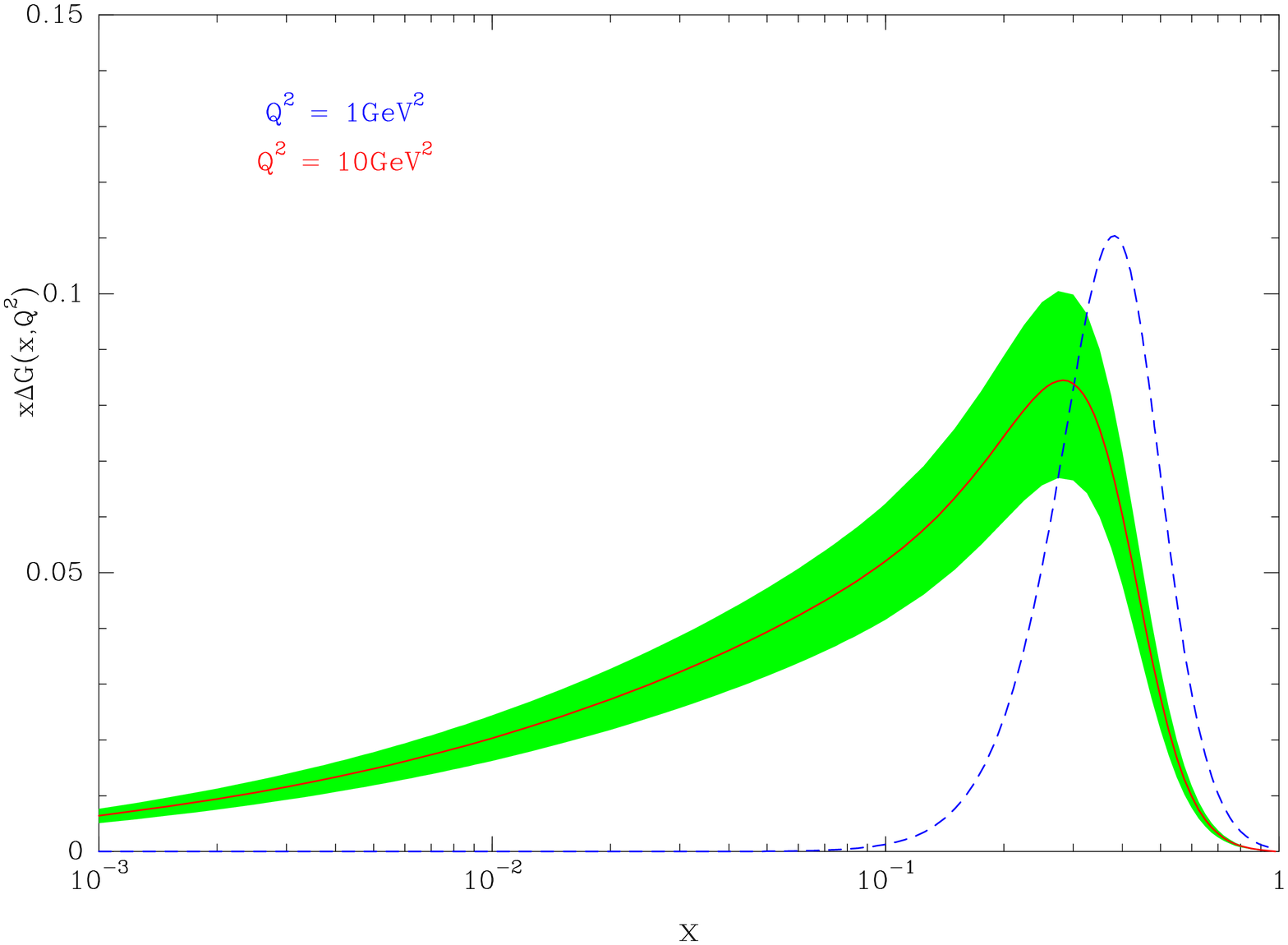}
  \includegraphics[width=8.5cm]{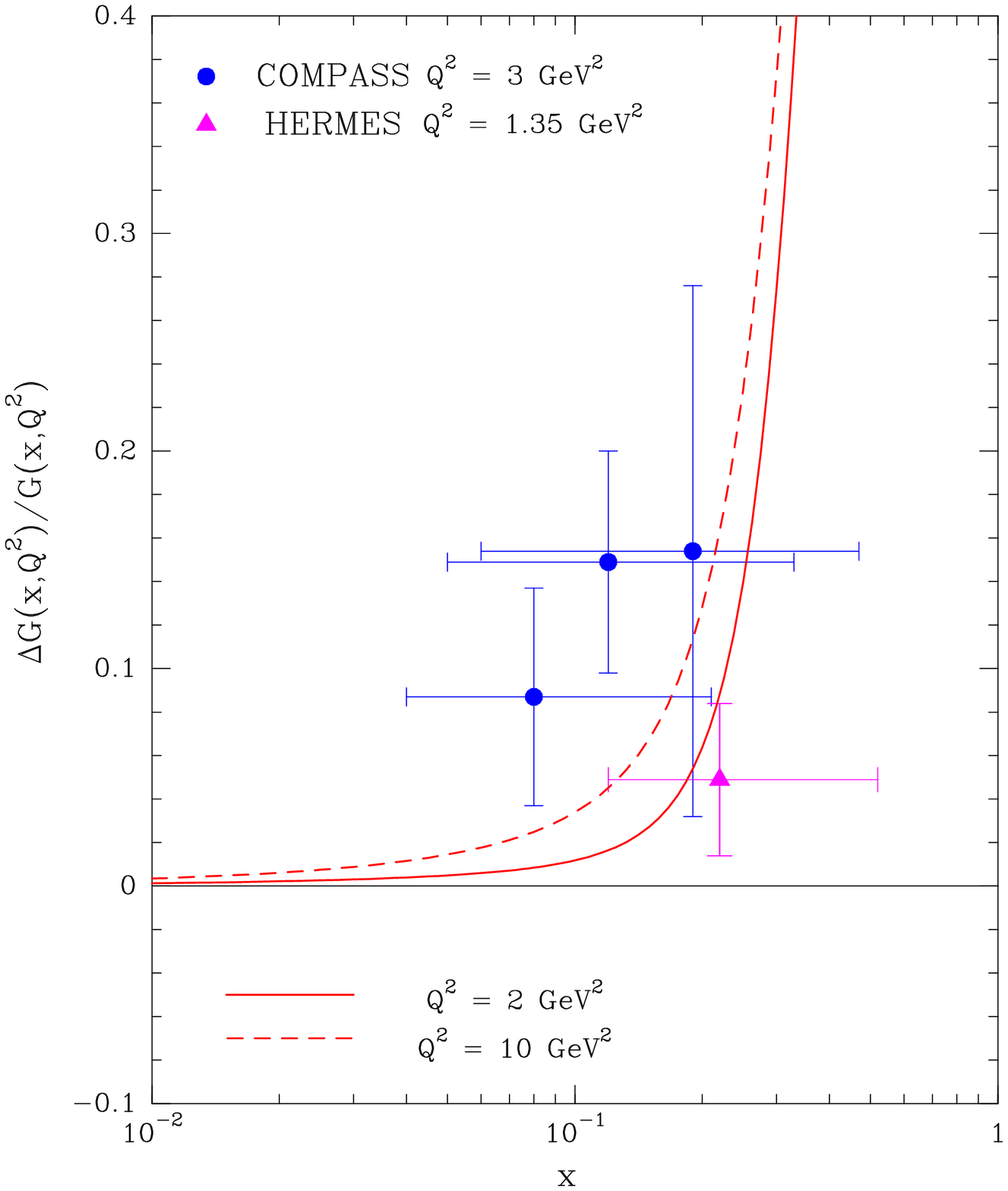}
  \caption[*]{\baselineskip 1pt
 {\it Top}~: The gluon helicity distribution $x\Delta G(x,Q^2)$ versus $x$,
for
$Q^2=1~\mbox{GeV}^2$ (dashed curve) and after NLO QCD evolution for
$Q^2=10~\mbox{GeV}^2$ (solid curve), with the corresponding error band  .\\
{\it Bottom}~: $\Delta G(x,Q^2)/G(x,Q^2)$ versus $x$, for $Q^2=2~\mbox{GeV}^2$
(solid curve) and $Q^2=10~\mbox{GeV}^2$ (dashed curve). The data are from
HERMES \cite{hermes} and COMPASS \cite{compass}}
\label{fgluon}
\end{center}
\end{figure}

By comparing with the results of 2002 \cite{bbs1}, we have observed a
remarkable stability of some important parameters, the light quarks potentials
$X_{0u}^{\pm}$ and  $X_{0d}^{\pm}$, whose numerical values are almost
unchanged
\footnote{Note the interesting relation $X_{0u}^- \simeq X_{0d}^-$, already
found in Ref. \cite{bbs1}.}. The new temperature is slightly lower. As a
result
the main features of the new light quark and antiquark distributions are only
scarcely modified. However it is instructive to note that in Tables 1 and 2, one
can judge the improvement obtained in this new version compared to the old 2002
version. The last column gives the $\chi^2$ obtained with the old
parameters and the new data set.\\
We display in Fig.\ref{unpolglobal} the different unpolarized parton
distributions $xf(x,Q^2)$ ($f=u, d, s, c, \bar u, \bar d, \bar s ~ \mbox{and}
{}~
 G$)  versus $x$, after NLO QCD evolution at $Q^2=10\mbox{GeV}^2$ , with the
corresponding error bands.  Similarly the different quark and antiquark
helicity distributions $x\Delta f(x,Q^2)$ ($f=u, d, s,\bar u, \bar d ~
\mbox{and} ~  \bar s $)  versus $x$, after NLO QCD evolution at
$Q^2=10\mbox{GeV}^2$ , with the corresponding error bands are shown in Fig.
\ref{polglobal}.\\
Our determination of the gluon helicity distribution deserves a special
discussion.  We display in Fig.~\ref{fgluon}{\it Top} the gluon helicity
distribution versus $x$ at the initial scale $Q_{0}^2 = 1\mbox{GeV}^2$ and
$Q^2
= 10\mbox{GeV}^2$. At the initial scale it is sharply peaked around $x=0.4$,
but this feature lessens after some QCD evolution. We note that $P(x)$
introduced above, has the following expression, $P(x)= 0.731
x^{5.210}/(x^{6.072} + 0.006)$, which is such that $0<P(x)<1$ for $0<x<1$, so
positivity is satisfied and the gluon helicity distribution remains positive.
As already mentioned the term $x^{d_G}$ plays an important role. It has a
strong effect on the quality of the fit of $g_1^{p,n,d}(x,Q¨^2)$, since the
$\chi^2$ increases substantially when $d_G$ decreases. Its value also affects
the shape of the gluon helicity distribution, which becomes larger towards the
smaller $x$-values, for smaller $d_G$.
We display $\Delta G(x,Q^2)/G(x,Q^2)$ in  Fig.~\ref{fgluon}{\it Bottom} for
two
$Q^2$ values and some data points \cite{hermes,compass}, which suggest that
the
gluon helicity distribution is positive indeed. According to the constraints
of
the counting rules this ratio should go to 1 when $x=1$, but we observe that
this is not the case here, since for example at the initial scale
$P(x=1)=0.726$. In some other parameterizations in the current literature,
this
ratio goes to zero, since the large $x$ behavior of $x \Delta G(x)$ is $(1 -
x)^{\beta}$, with $\beta >> 3$ \cite{dssv1,others}. Clearly one needs a better
knowledge of $\Delta G(x,Q^2)/G(x,Q^2)$ for $x > 0.2$.

\newpage

\section{Deep inelastic scattering}
\subsection{Unpolarized DIS experiments}
\begin{figure}[hbp]   
\begin{center}
\includegraphics[width=12.0cm]{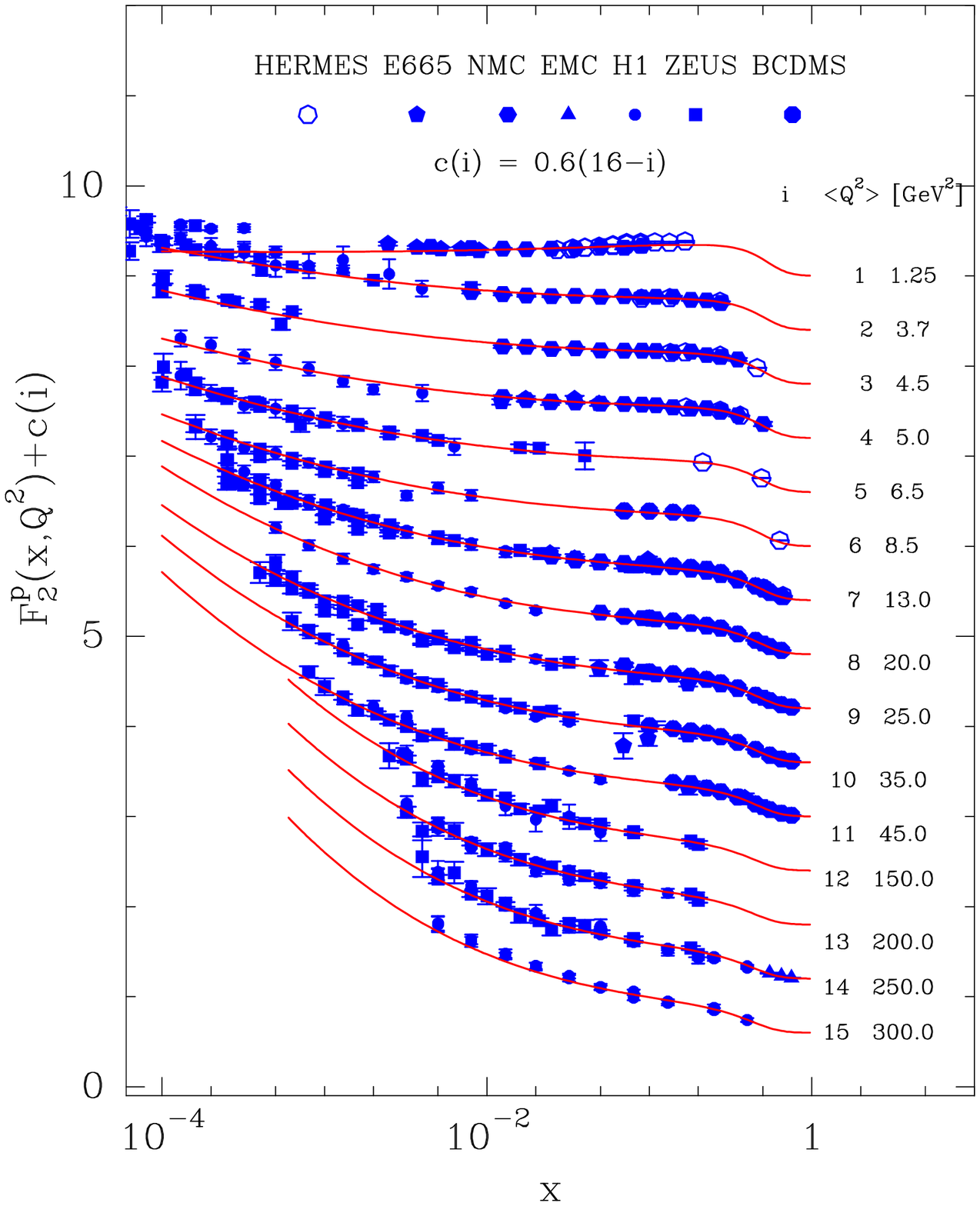}
\caption[*]{\baselineskip 1pt
 $F^{p}_{2}(x,Q^2)$ as a function of $x$ for fixed $\langle Q^2 \rangle$ and
data from HERMES \cite{hermes1}, E665 \cite{e66596}, NMC \cite{nmc97b}, EMC
\cite{emc}, H1 \cite{h109,h111}, ZEUS \cite{zeus01}, BCDMS \cite{bcdms}. The
function $c(i) = 0.6(16 - i )$, $i=1$ corresponds to $\langle Q^2 \rangle =
1.25 \mbox{GeV}^2$ . The curves are the results of the statistical approach.}
\label{f2p}
\end{center}
\end{figure}

\begin{figure}[hbp]   
\begin{center}
\includegraphics[width=12.0cm]{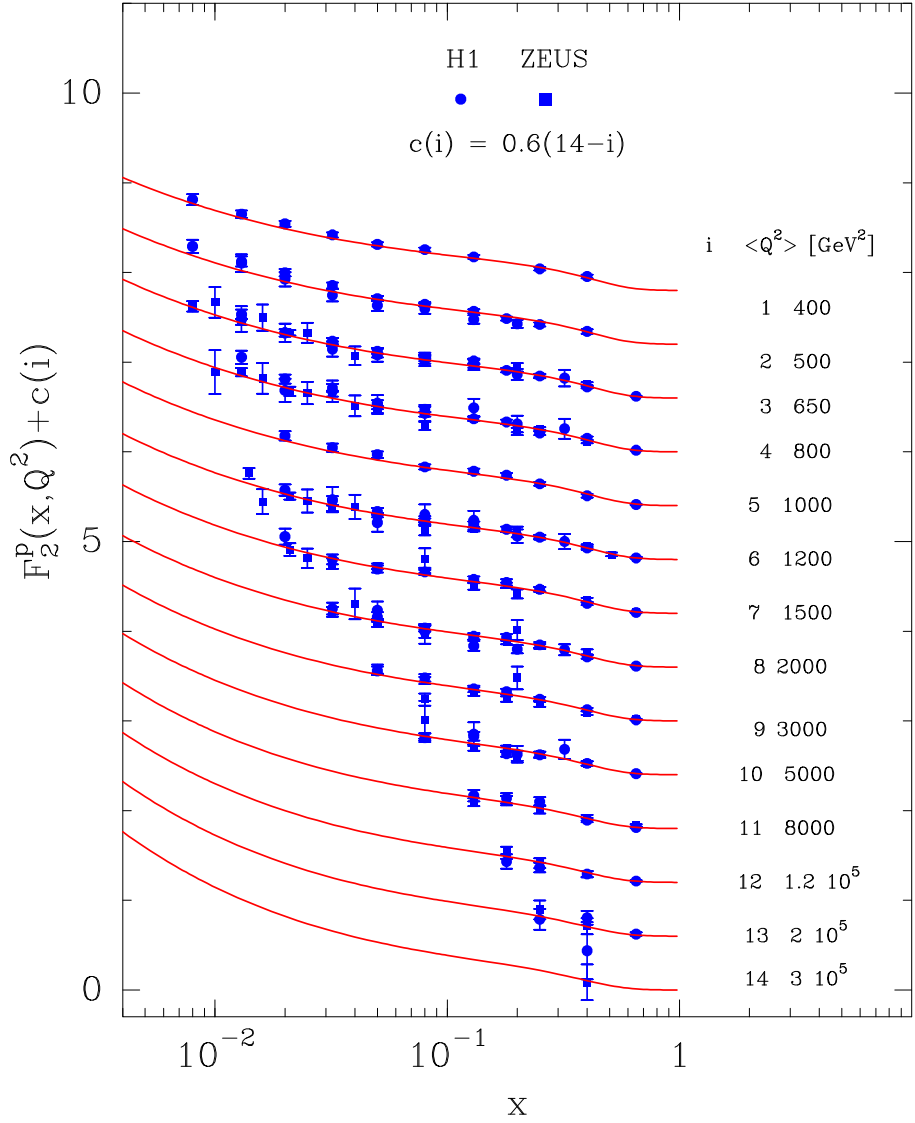}
\caption[*]{\baselineskip 1pt
 $F^{p}_{2}(x,Q^2)$ as a function of $x$ for fixed high $\langle Q^2 \rangle$
and data from  H1 \cite{h109,h111}, ZEUS \cite{zeus01}. The function $c(i) =
0.6(14 - i )$, $i=1$ corresponds to $\langle Q^2 \rangle = 400\mbox{GeV}^2$ .
The curves are the results of the statistical approach. }
\label{F2p}
\end{center}
\end{figure}

First we present some selected experimental tests for the unpolarized PDF by
considering $\mu N$ and $e N$ DIS, for which several experiments have yielded
a
large number of data points on the structure functions $F_2^N(x,Q^2)$, $N$
stands for either a proton or a deuterium target. We have used fixed target
measurements which probe a rather limited kinematic region in $Q^2$ and $x$
and
also HERA data which cover a very large $Q^2$ range and probe the very low $x$
region, dominated by a fast rising  behavior, consistent with our diffractive
term (See Eq. (\ref {eq1})).\\
For illustration of the quality of our fit and, as an example, we show in
Fig.~\ref{f2p} and Fig.~\ref{F2p}, our results for $F_2 ^p (x,Q^2)$ on
different fixed proton targets, together with H1 and ZEUS data . We note that
the analysis of the scaling violations leads to a gluon distribution
$xG(x,Q^2)$, in fairly good agreement with our simple parametrization (See Eq.
(\ref{eq3})). 

\begin{figure}[hbp]  
\begin{center}
\includegraphics[width=12.0cm]{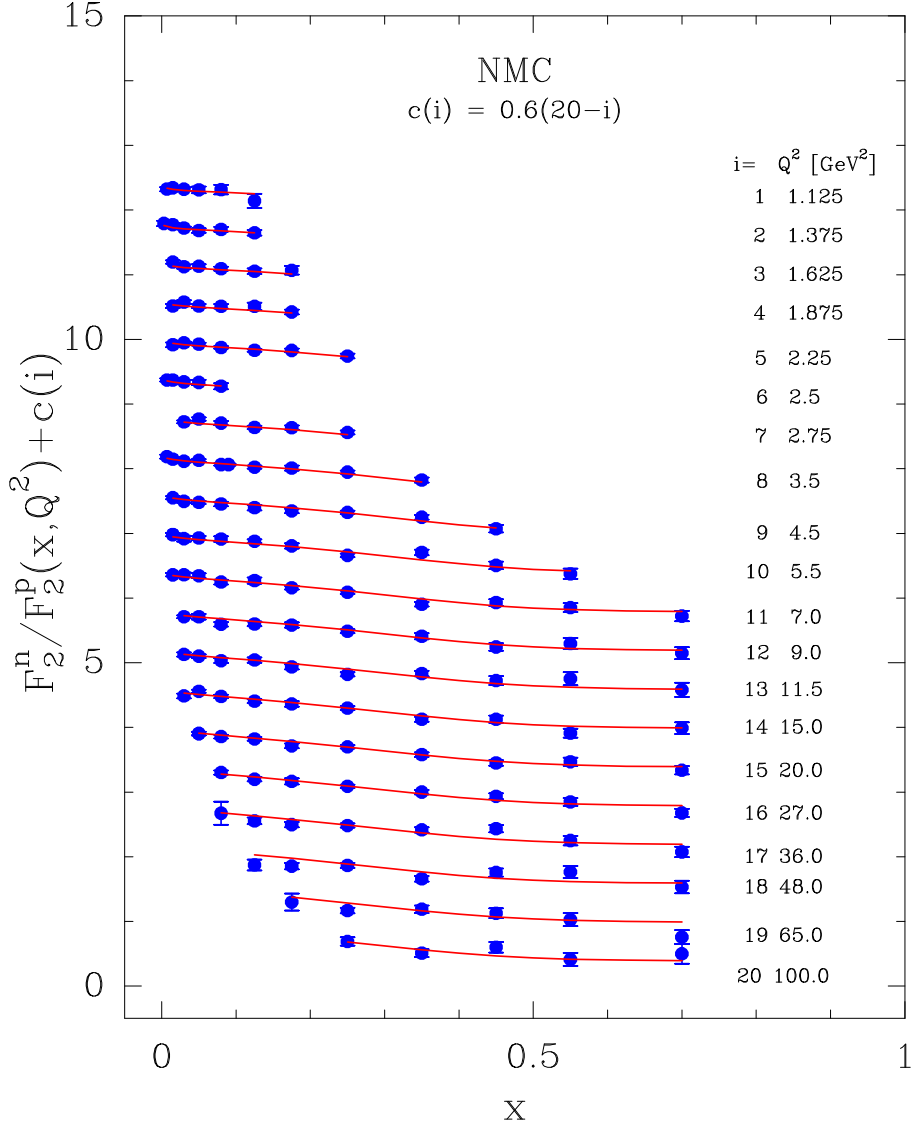}
\caption[*]{\baselineskip 1pt
$F^{n}_{2}/F^{p}_{2}(x,Q^2)$ as a function of $x$ for fixed $\langle Q^2
\rangle$ and data from NMC \cite{nmc1}. The function $c(i) = 0.6(20 - i )$,
$i=1$ corresponds to $\langle Q^2 \rangle = 1.125 \mbox{GeV}^2$. The curves
are
the results of the statistical approach.}
\label{nmc}
\end{center}
\end{figure}

\begin{figure}[hbp]   
\begin{center}
\includegraphics[width=12.0cm]{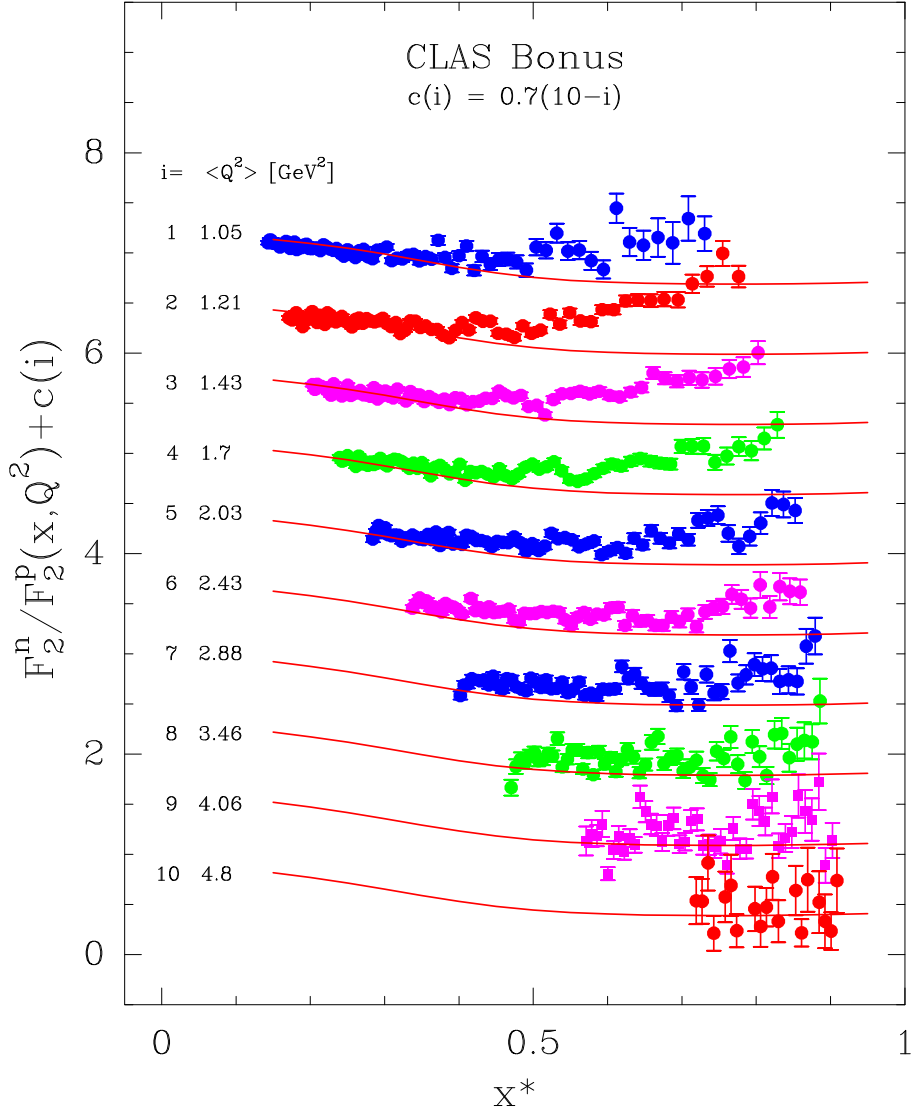}
\caption[*]{\baselineskip 1pt
 $F^{n}_{2}/F^{p}_{2}(x,Q^2)$ as a function of $x^* \simeq x$ (for the exact
definition of $x^*$ see Ref. \cite{bonus}) for fixed $\langle Q^2
\rangle$ and data from CLAS BoNus \cite{bonus}. The function $c(i) = 0.7(10 -
i
)$, $i=1$ corresponds to $\langle Q^2 \rangle = 1.05 \mbox{GeV}^2$. The curves
are the results of the statistical approach}
\label{bonus}
\end{center}
\end{figure}

Another rather interesting physical quantity is the neutron $F_2 ^n$ structure
function and in particular the ratio $F_2 ^n/F_2 ^p (x,Q^2)$ which provides
strong contraints on the PDF of the nucleon. For example the behavior of this
ratio at
large $x$ is directly related to the ratio of the $d$ to $u$ quarks in the
limit $x \to 1$, a long-standing problem for the proton structure. We show the
results of two experiments, NMC in Fig.~\ref{nmc}, which is very accurate and
covers a reasonable $Q^2$ range up to $x=0.7$ and CLAS  in Fig. ~\ref{bonus},
which covers a smaller $Q^2$ range up to larger $x$ values, both are fairly
well described by the statistical approach. Several comments are in order. In
the small $x$ region this ratio, for both cases, tends to 1 because the
structure functions are dominated by sea quarks driven by our universal
diffractive term. In the high $x$ region dominated by valence quarks, the NMC
data suggest that this ratio goes to a value of the order of 0.4 for $x$ near
1, which corresponds to the value 0.16 for $d(x)/u(x)$ when $x \to 1$, as
found
in the statistical approach \cite{bbs4}. The CLAS data at large $x$ cover the
resonance region of the cross section and an important question is whether
Bloom-Gilman duality holds as well for the neutron as it does for the proton.
We notice that the predictions of the statistical approach suggest an
approximate validity of this duality, except for some low $Q^2$ values.
A better precision and the extension of this experiment with the 12GeV
Jefferson Lab will certainly provide even stronger constraints on PDFs up to
$x
\simeq 0.8$.

\begin{figure}[hbp]  
\vspace*{-15.5ex}
\begin{center}
 \includegraphics[width=7.5cm]{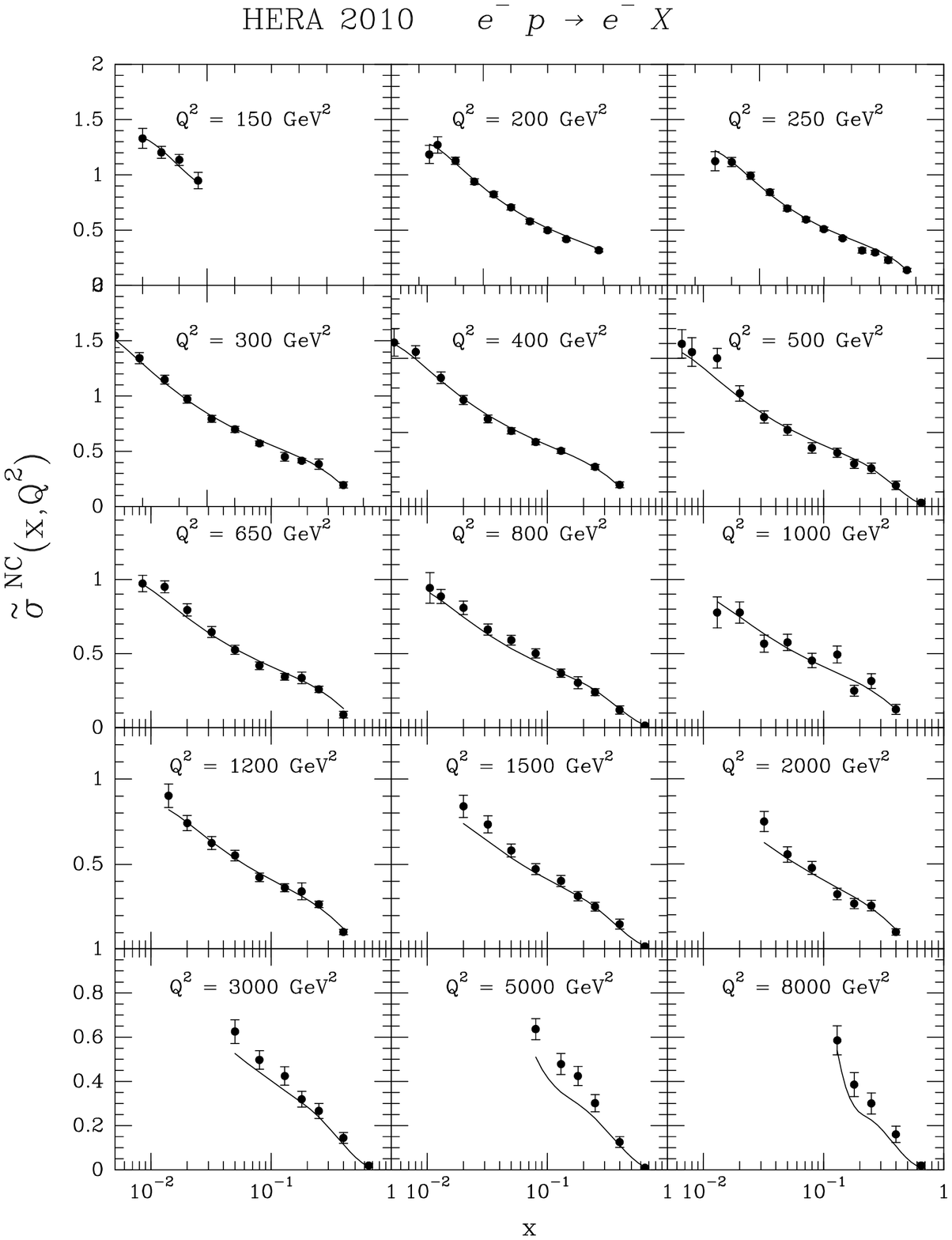}
  \includegraphics[width=7.5cm]{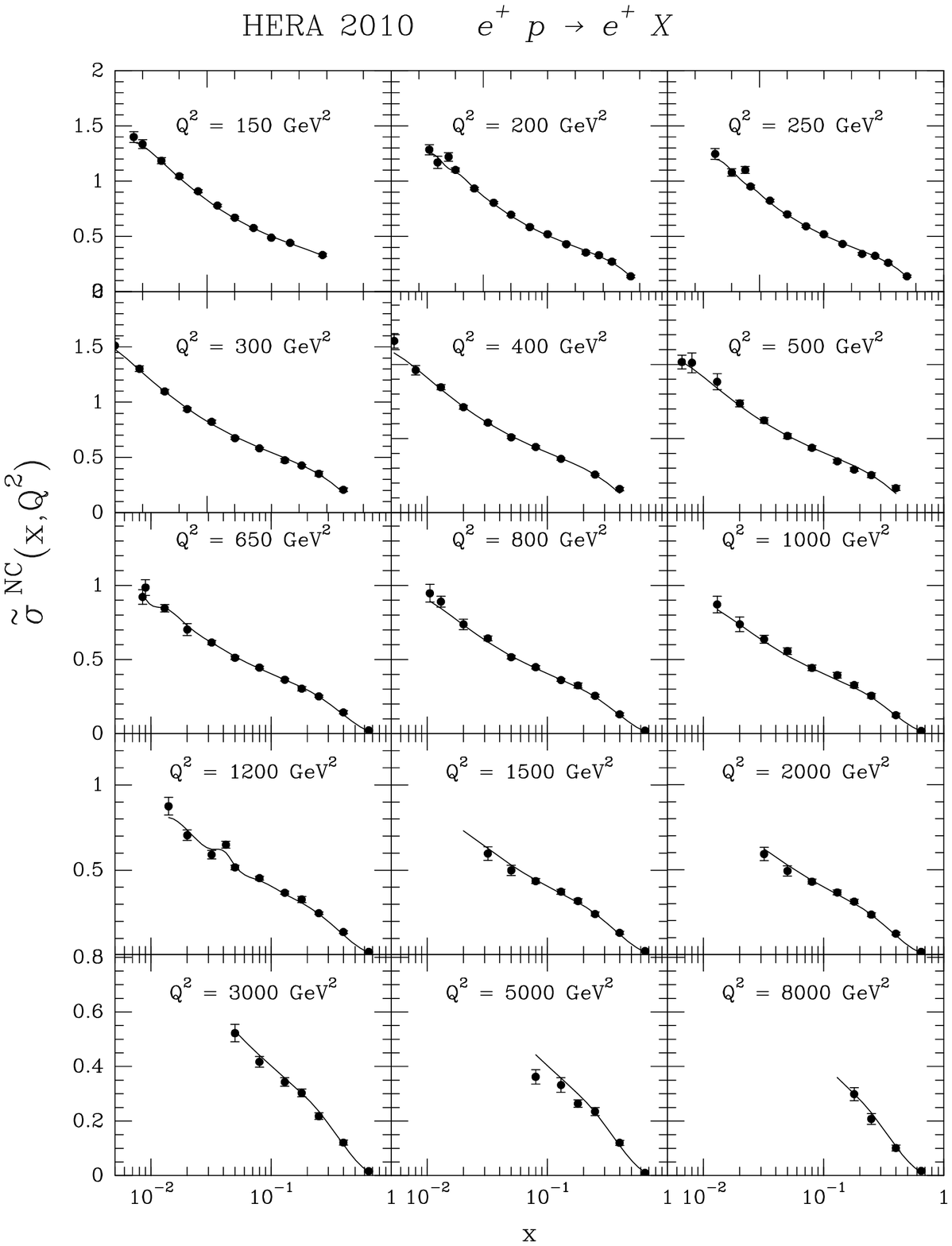}
  \caption[*]{\baselineskip 1pt
 {\it Top}~:  Comparison of the data on the reduced neutral cross section
$\tilde \sigma^{NC}(x,Q^2)$, in $e^- p$ collisions as a function of $x$ and
for
different $Q^2$ values, with the results of the statistical approach. Data are
from HERA \cite{hera10}.\\
{\it Bottom}~: Same for $e^+ p$ collisions.}
\label{NC}
\end{center}
\end{figure}

\begin{figure}[hbp]   
\begin{center}
\includegraphics[width=10.0cm]{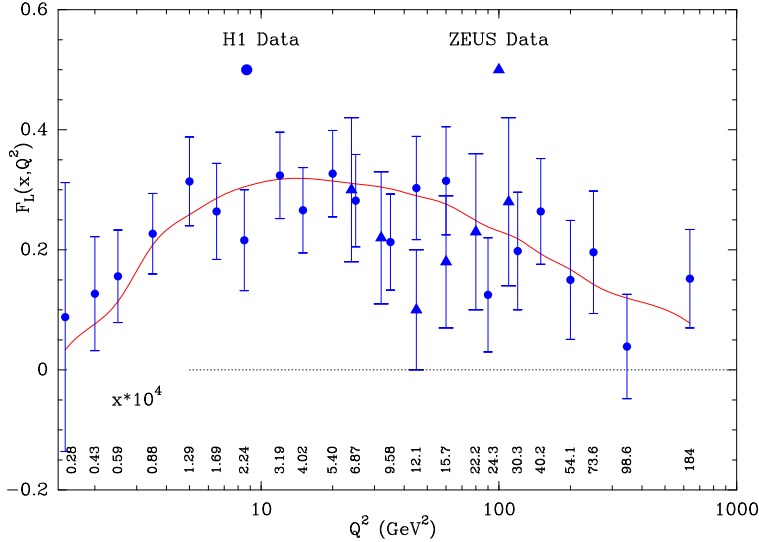}
\caption[*]{\baselineskip 1pt
The longitudinal proton structure function $F_L(x,Q^2)$ averaged in $x$ at
given $Q^2$ values. Data are from ZEUS \cite{zeus09} and H1 \cite{h114} and
the
curve is the result of the statistical approach.}
\label{FL}
\end{center}
\vspace*{-3.5ex}
\end{figure}

\begin{figure}[htbp]  
\begin{center}
\includegraphics[width=10.0cm]{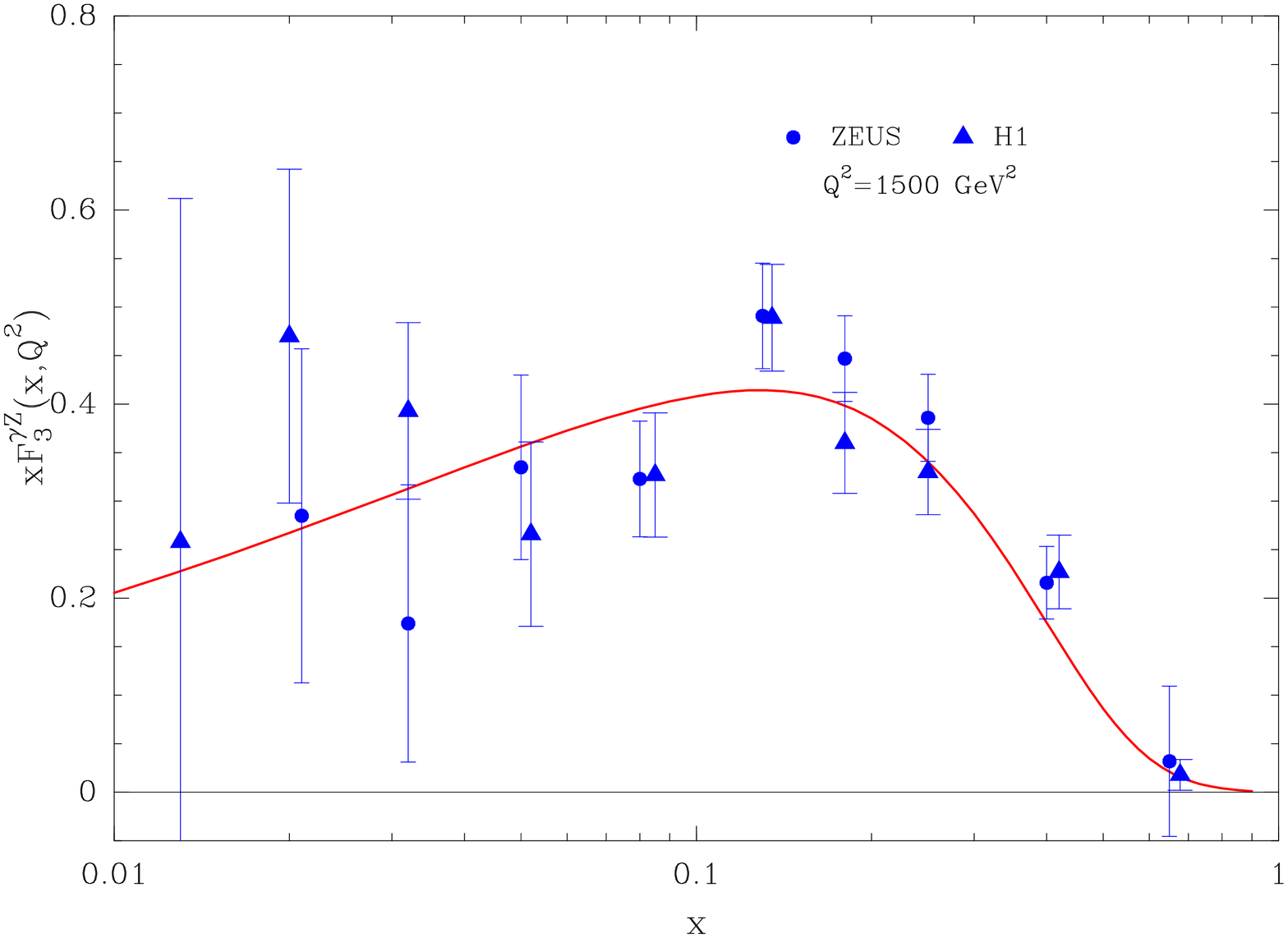}
\caption[*]{\baselineskip 1pt
 The interference term $xF_3^{\gamma Z}$ extracted in $e^{\pm} p$ collisions
at
HERA. Data are from ZEUS \cite{zeus13}  and H1 \cite{h112}, the curve is the
prediction of the statistical approach.}
\label{FZ}
\end{center}
\end{figure}

\clearpage
\newpage
\vspace*{-15.5ex}
We now turn to the inclusive neutral and charged current $e^{\pm} p$ cross
sections which, in addition to $F_2^p$, give access to other structure
functions.\\
The neutral current DIS processes have been measured at HERA in a kinematic 
region where both the $\gamma$ and the $Z$ exchanges must be considered.
The cross sections for neutral current can be written, at lowest order, as 
\begin{equation}
\frac{d^2 \sigma^{\pm}_{NC}}{dx dQ^2} =
\frac{2\pi \alpha^2}{x Q^4}
\left[Y_+ \tilde F_2(x, Q^2) \mp Y_- x \tilde F_3(x, Q^2) - y^2 F_L(x,
Q^2)\right]~,
\label{dcrossnc}
\end{equation}
where
\begin{equation}
\tilde F_2(x, Q^2)\!\! =\!\! F_2^{\gamma}(x,Q^2) -v_e\chi_z(Q^2)F_2^{\gamma
Z}(x,Q^2)+\!\!
(a_e^2 + v_e^2)\chi^2_z(Q^2) F_2^{Z} (x, Q^2)~,
\label{tildf2}
\end{equation}
\begin{equation}
x\tilde F_3(x, Q^2) = -a_e\chi_z(Q^2) xF_3^{\gamma Z}(x, Q^2) + 2a_e v_e
\chi^2_z(Q^2)
xF_3^Z (x, Q^2)~.
\label{tildf3}
\end{equation}
The structure function $F_L(x, Q^2)$ is sizeable only at high $y$ and we will
come back to it later.
The other structure functions introduced above, have the following 
expressions in terms of the parton distributions \footnote{ For simplicity we
write them at LO, but the calculations were done by including the NLO
corrections \cite{mover}.}
\begin{eqnarray}
\left[F_2^{\gamma}, F_2^{\gamma Z}, F_2^{ Z}\right](x, Q^2) &=&
\sum_f \left[e^2_f, 2e_f v_f, a^2_f + v^2_f\right]
\left(xq_f(x, Q^2) + x\bar q_f(x, Q^2)\right) , \nonumber \\
\left[xF_3^{\gamma Z}, xF_3^Z \right](x, Q^2) &=& \sum_f \left[2e_f a_f, 2a_f
v_f\right]
\left(xq_f(x, Q^2) - x \bar q_f(x, Q^2)\right) .
\label{fgh}
\end{eqnarray}
Here the kinematic variables are
$y = Q^2/xs$, $Y_{\pm} = 1 \pm (1-y)^2$, $\sqrt{s} = 2\sqrt{E_e E_p}$,
$E_e$ and $E_p$ are the electron (positron) and proton beam energies
respectively. Morever, $v_i$ and $a_i$ are the vector and axial-vector
weak coupling constants for the lepton $e$ and the quark $f$, respectively,
and $e_f$ is the charge. The function $\chi_z(Q^2)$ is given by
\begin{equation}
\chi_z(Q^2) = \frac{1}{\sin^2{2\theta_W}}
\frac{Q^2}{Q^2 + M^2_Z}~,
\label{fchi2}
\end{equation}
where $\theta_W$ is the weak mixing angle and $M_Z$ is the $Z$-boson mass.
The reduced cross sections are defined as
\begin{equation}
\tilde \sigma^{\pm}_{NC}(x, Q^2) = 
\frac{Q^4 x}{Y_+ 2\pi \alpha^2}
\frac{d^2 \sigma^{\pm}_{NC}}{dx dQ^2}~.
\label{redncross}
\end{equation}
Our predictions are compared with HERA data in 
Fig.~\ref{NC}, as a function of $x$, 
in a broad range of $Q^2$ values and the agreement is good.\\
For low $Q^2$, the contribution of the longitudinal structure function
$F_L(x,Q^2)$.  to the cross section at HERA is only sizeable at $x$ smaller
than approximately $10^{-3}$ and in this domain the gluon density dominates
over the sea quark density. More precisely, it was shown that \cite{am78}
\begin{eqnarray}
F_L(x,Q^2) &=& \frac{\alpha_s(Q^2)}{\pi}\left[\frac{4}{3}
\int_x^1 \frac{dy}{y}(\frac{x}{y})^2 F_2(y,Q^2)\nonumber \right. \\
&&\left. +2\Sigma_i e_i^2 \int_x^1  \frac{dy}{y}(\frac{x}{y})^2
(1 -\frac{x}{y})y G(y,Q^2)\right]
\end{eqnarray}
Before HERA was shut down, a dedicated run period, with reduced proton beam
energy, was
approved, allowing H1 to collect new results on $F_L$. We show on Fig.
\ref{FL}
 the expectations
of the statistical approach compared to the new data, whose precision is
reasonable. The trend and the magnitude of the prediction are in fair
agreement
with the data, so this is another test of the good predictive power of our
theoretical framework.
One can also test the behavior of the interference term between the photon and
the $Z$ exchanges, which can be isolated in neutral current $e^{\pm}p$
collisions at high $Q^2$. We have to a good approximation, if sea quarks are
ignored, $xF_3^{\gamma Z}(x,Q^2) = \frac{x}{3}(2u_v + d_v)(x,Q^2)$ and
comparison between data and prediction is shown in Fig. \ref{FZ}.\\

\begin{figure}[htbp]  
\begin{center}
\hspace*{0.5ex}
 \includegraphics[width=6.5cm]{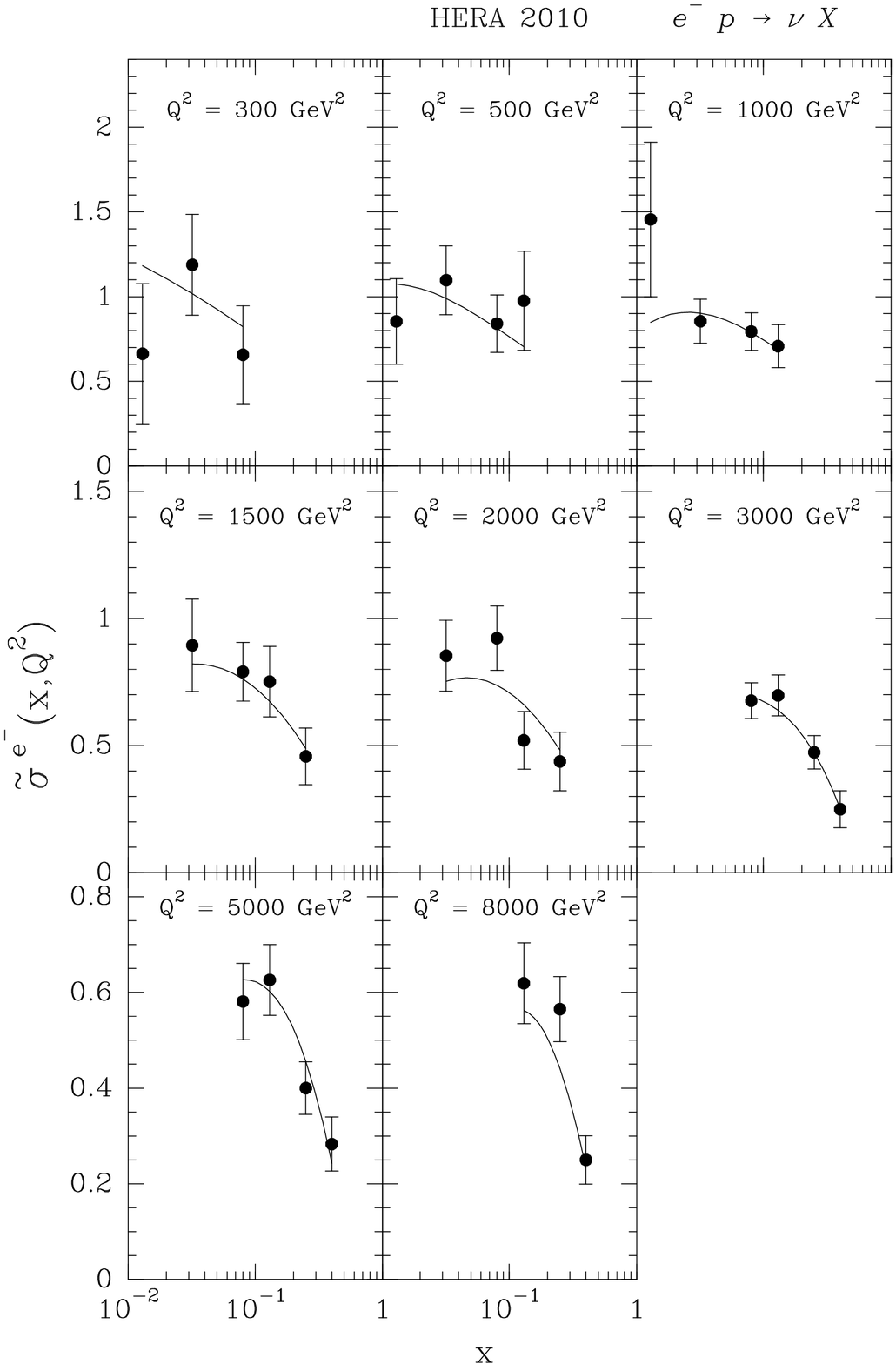}
  \includegraphics[width=6.5cm]{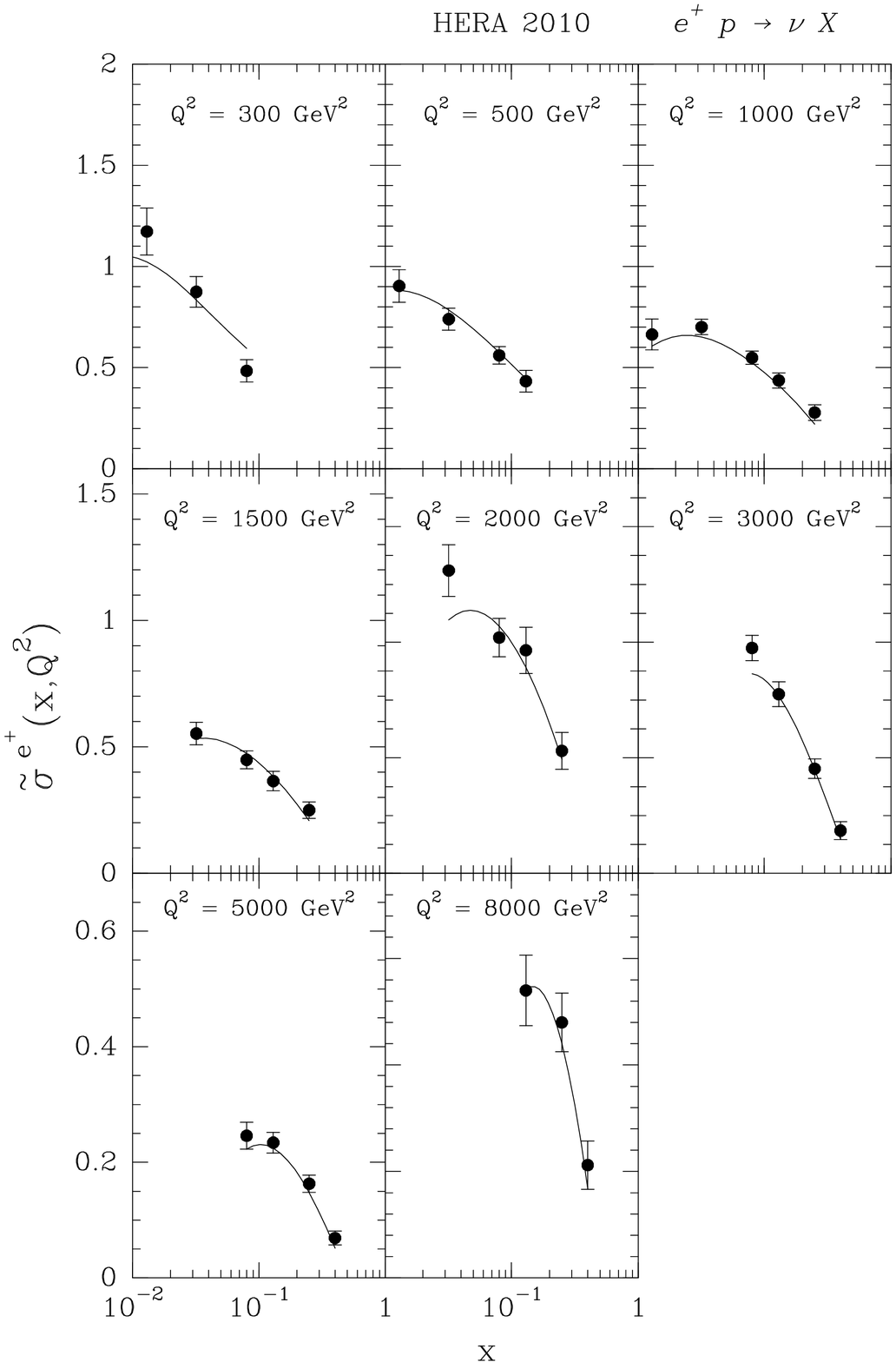}
  \caption[*]{\baselineskip 1pt
 {\it Left}~: Comparison of the data on the reduced charged cross section
$\tilde \sigma^{e^{-}}(x,Q^2)$, in $e^- p$ collisions as a function of $x$ and
for
different $Q^2$ values, with the results of the statistical approach. Data are
from HERA \cite{hera10}.\\
{\it Right}~: Same for $e^+ p$ collisions.}
\label{cc}
\end{center}
\end{figure}

\begin{figure}[hbp]    
\vspace*{-19.5ex}
\begin{center}
 \includegraphics[width=7.0cm]{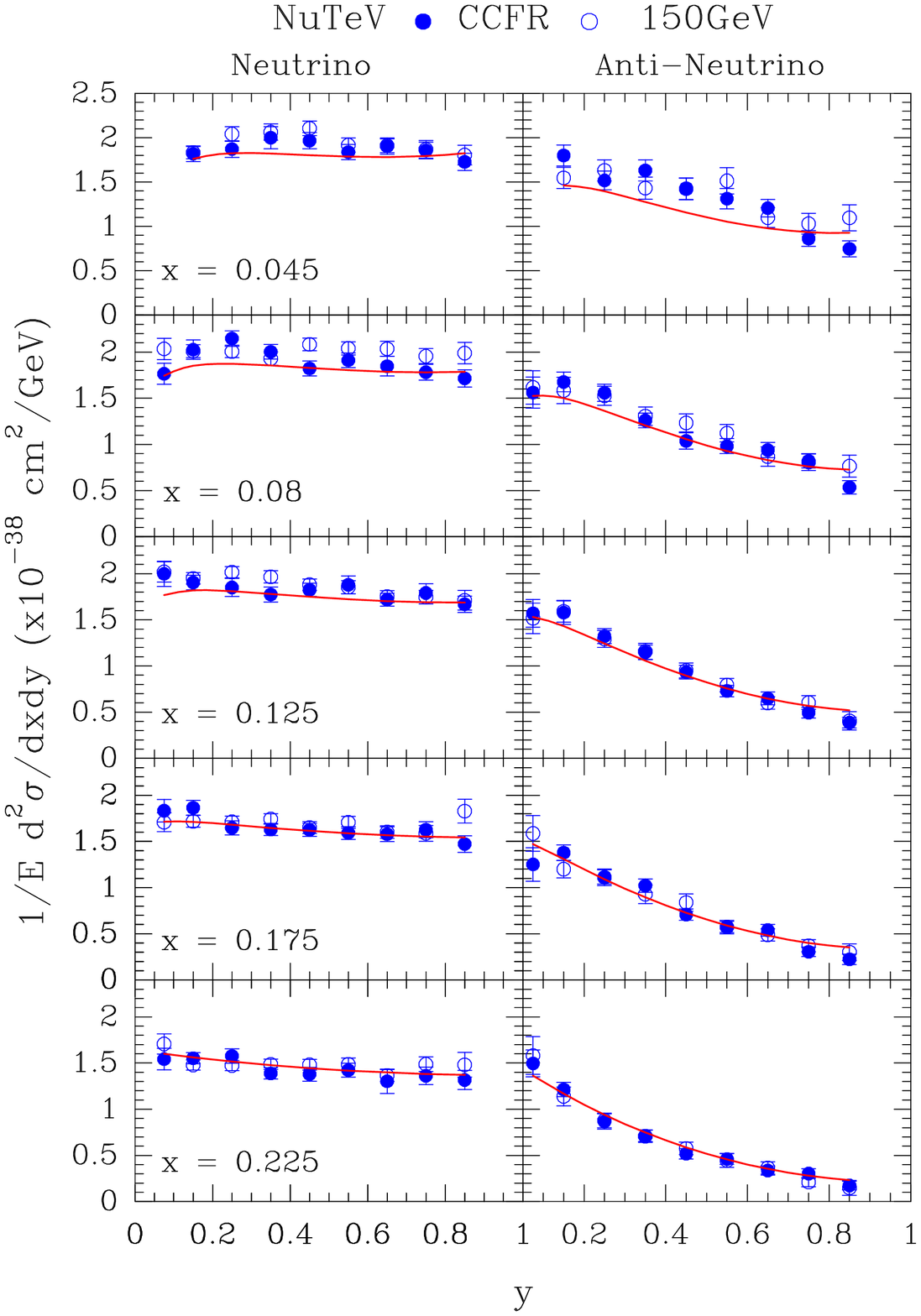}
  \includegraphics[width=7.0cm]{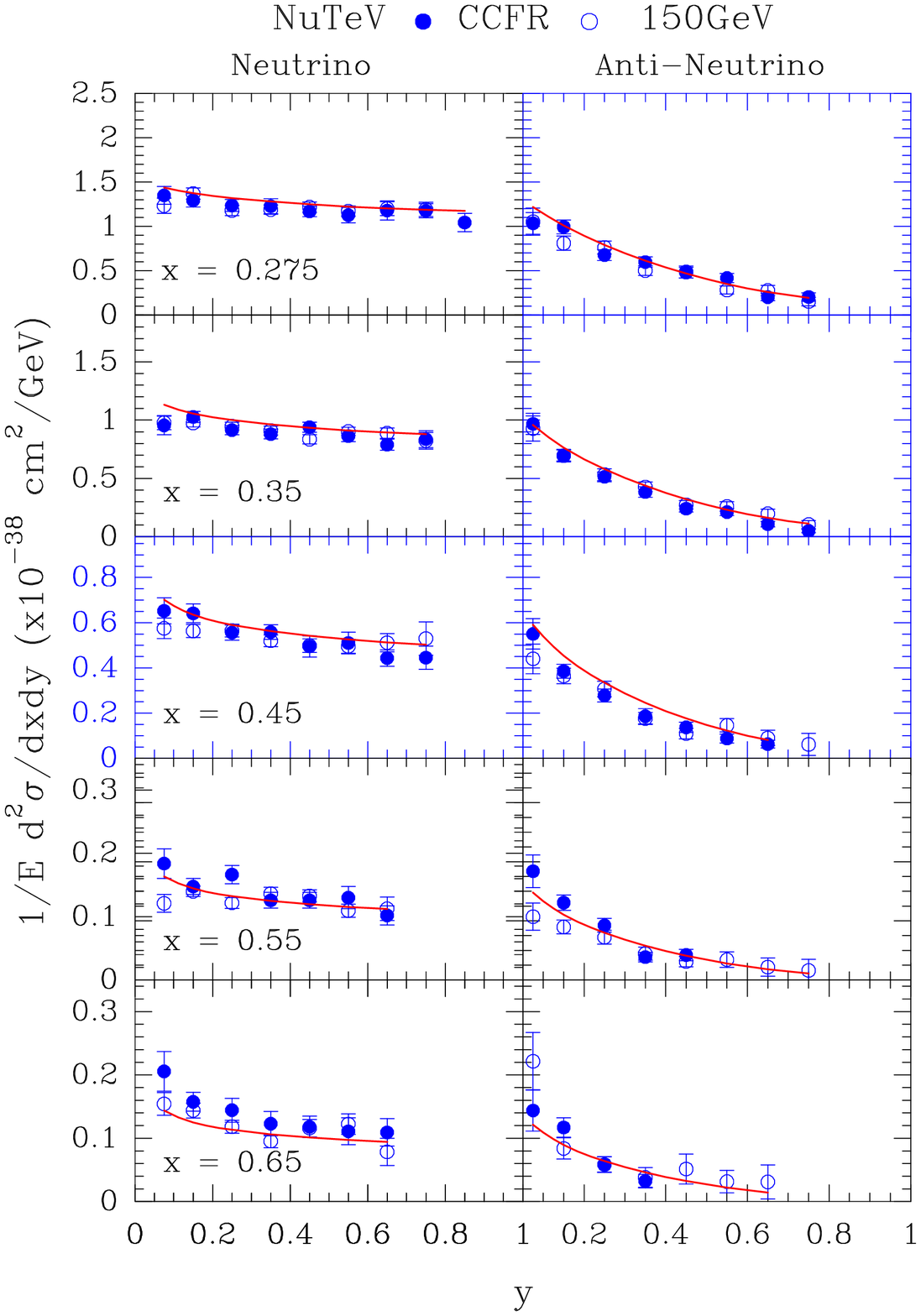}
  \caption[*]{\baselineskip 1pt
  Comparison of the data on the differential cross sections $\nu ({\bar \nu})
N$ for $E_\nu$ = 150 GeV, as a function of $y$ and for different $x$ values,
with the results of the statistical approach. Data are from CCFR \cite{ccfr01}
and NuTeV \cite{nutev06}.}
\label{nuanu}
\end{center}
\end{figure}

The charged current DIS processes have been also measured accurately at HERA
in an extented kinematic region. It has a serious impact on the determination
of the unpolarized parton distributions by allowing a flavor separation
because
they involve only the $W^{\pm}$ exchange. The cross sections are expressed, at
lowest order, in terms of three structure functions as follows 
\begin{eqnarray}
\frac{d^2 \sigma^{e^{\pm}}_{Born}}{dx dQ^2} &=&
\frac{G_F^2}{4\pi}\frac{M^4_W}{(Q^2 + M^2_W)^2}
\left[Y_+ F^{cc}_2 (x, Q^2) - y^2 F^{cc}_L(x, Q^2) \right. \nonumber \\
&&\left. {\mp} Y_- xF^{cc}_3(x, Q^2) \right]~,
\label{chargcur}
\end{eqnarray}
and the reduced cross sections are defined as
\begin{equation}
\tilde \sigma^{e^{\pm}}(x, Q^2) = 
\left[\frac{G_F^2}{4\pi}\frac{M^4_W}{(Q^2 + M^2_W)^2} \right]^{-1}
\frac{d^2 \sigma^{cc}}{dx dQ^2}~.
\label{redcross}
\end{equation}
At leading order for $e^- p \rightarrow \nu_e X$ with a longitudinally 
polarized beam
\begin{eqnarray}
F^{cc}_2(x, Q^2) &=& x[u(x, Q^2) + c(x, Q^2) + \bar d(x, Q^2) +
\bar s (x, Q^2)] \nonumber \\
xF^{cc}_3(x, Q^2) &=& x[u(x, Q^2) + c(x, Q^2) - \bar d(x, Q^2)
-\bar s(x, Q^2)]~,
\label{f2em}
\end{eqnarray}
and for $e^+ p \rightarrow \bar \nu_e X$
\begin{eqnarray}
F^{cc}_2(x, Q^2) &=& x[d(x, Q^2) + s(x, Q^2) + \bar u(x, Q^2) +
\bar c (x, Q^2)] \nonumber \\
xF^{cc}_3(x, Q^2) &=& x[d(x, Q^2) + s(x, Q^2) - \bar u(x, Q^2)
-\bar c(x, Q^2)]~.
\label{f2ep}
\end{eqnarray}
At NLO in QCD $F^{cc}_L$ is non zero, but it gives negligible contribution, 
except at $y$ values close to 1.
Our predictions for $\sigma^{cc}(x, Q^2)$ at NLO are compared with H1 and ZEUS
data in Fig.~\ref{cc}, as a function of 
$x$ in a broad range of $Q^2$ values.\\
The differential inclusive neutrino and antineutrino cross sections have the 
following standard expressions
\begin{eqnarray}
\frac{d^2 \sigma^{\nu, (\bar \nu)}}{dx dy} &=&
\frac{G^2_F M_p E_{\nu}}{ \pi (1 + \frac{Q^2}{M^2_W})^2}
\left[x y^2 F_1^{\nu (\bar \nu)}(x, Q^2)
+ (1-y-\frac{M_p xy}{2E_{\nu}})
F_2^{\nu (\bar \nu)}(x, Q^2) \right. \nonumber \\
&& \left. \pm (y - \frac{y^2}{2})x F_3^{\nu (\bar \nu)}(x, Q^2)\right]\,,
\label{dcrossnu}
\end{eqnarray}
$y$ is the fraction of total leptonic energy transfered to the 
hadronic system and $E_{\nu}$ is the incident neutrino  energy.
$F_2$ and $F_3$ are given by
Eq.~(\ref{f2em}) for $\nu p$ and Eq.~(\ref{f2ep}) for $\bar \nu p$, and $F_1$
is related to $F_2$ by
\begin{equation}
2 x F_1 = \frac{1 + 4 x^2 M^2_{p}/Q^2}{1 + R} F_2\,,
\label{f1vf2}
\end{equation}
where $R = \sigma_L / \sigma_T$, the ratio of the longitudinal to
transverse cross sections of the W-boson production. Our results at NLO
compared with the CCFR and NuTeV
data are shown in Fig.~\ref{nuanu}. As
expected,
for fixed $x$, the $y$ 
dependence is rather flat for neutrino and has the characteristic 
$(1-y)^2$ behavior for antineutrino.\\ 

\newpage

\subsection{Polarized DIS experiments}

The spin-dependent structure function  $g_1(x, Q^2)$ has the well-known NLO
QCD expression \cite{wv}
\begin{eqnarray}
g_1(x,Q^2)&=&\frac{1}{2}\sum _{q} ^{n_f}e_{q}^2
[(\Delta q +\Delta\bar{q})\otimes (1 +
\frac{\alpha_s(Q^2)}{2\pi}\delta C_q)
\nonumber\\
&&+\frac{\alpha_s(Q^2)}{2\pi}\Delta G\otimes \frac{\delta C_G}
{n_f}], \label{g1}
\end{eqnarray}
$\Delta q(x,Q^2), \Delta\bar{q}(x,Q^2)$ and $\Delta G(x,Q^2)$ are
quark, antiquark and gluon helicity distributions in the nucleon. $\delta
C(x)_{q,G}$ are the NLO spin-dependent Wilson
coefficient functions and the symbol $\otimes$ denotes the usual
convolution in Bjorken $x$ space. $n_f$ is the number of active
flavors for light quarks. \\
We recall that according to the results shown in Section 3, we have obtained a
good flavor separation of these helicity distributions: for all $x$ and $Q^2$
values, $\Delta u > 0$ is the largest one, $\Delta d < 0$ is smaller in
magnitude, $\Delta \bar u > 0$ and $\Delta \bar d  < 0$ are approximately
opposite and  $\Delta s < 0$, $\Delta \bar s < 0$ are much smaller.\\

\begin{figure}[hbp]   
\vspace*{-10.5ex}
\begin{center}
 \includegraphics[width=12.5cm]{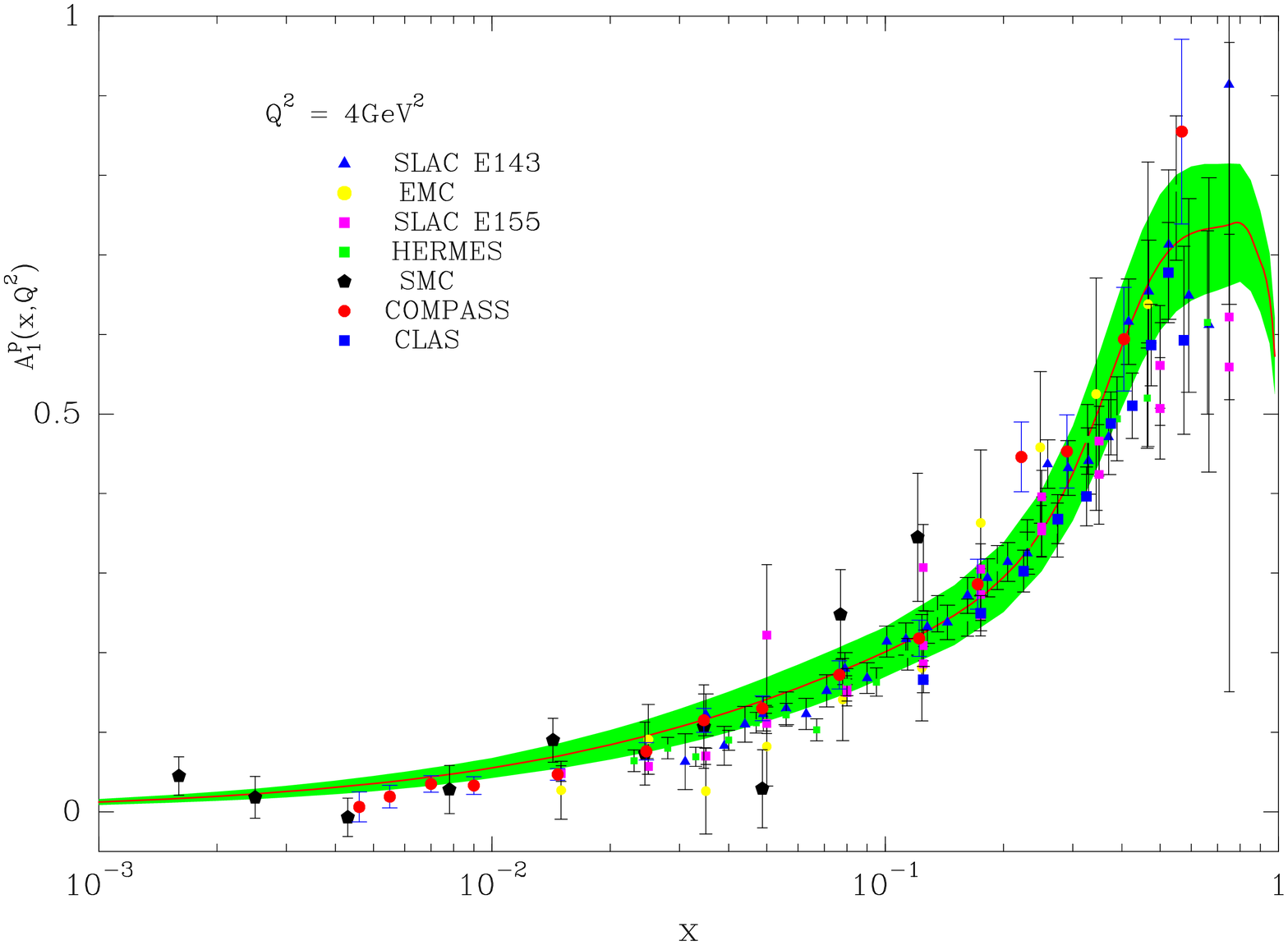}
  \includegraphics[width=12.5cm]{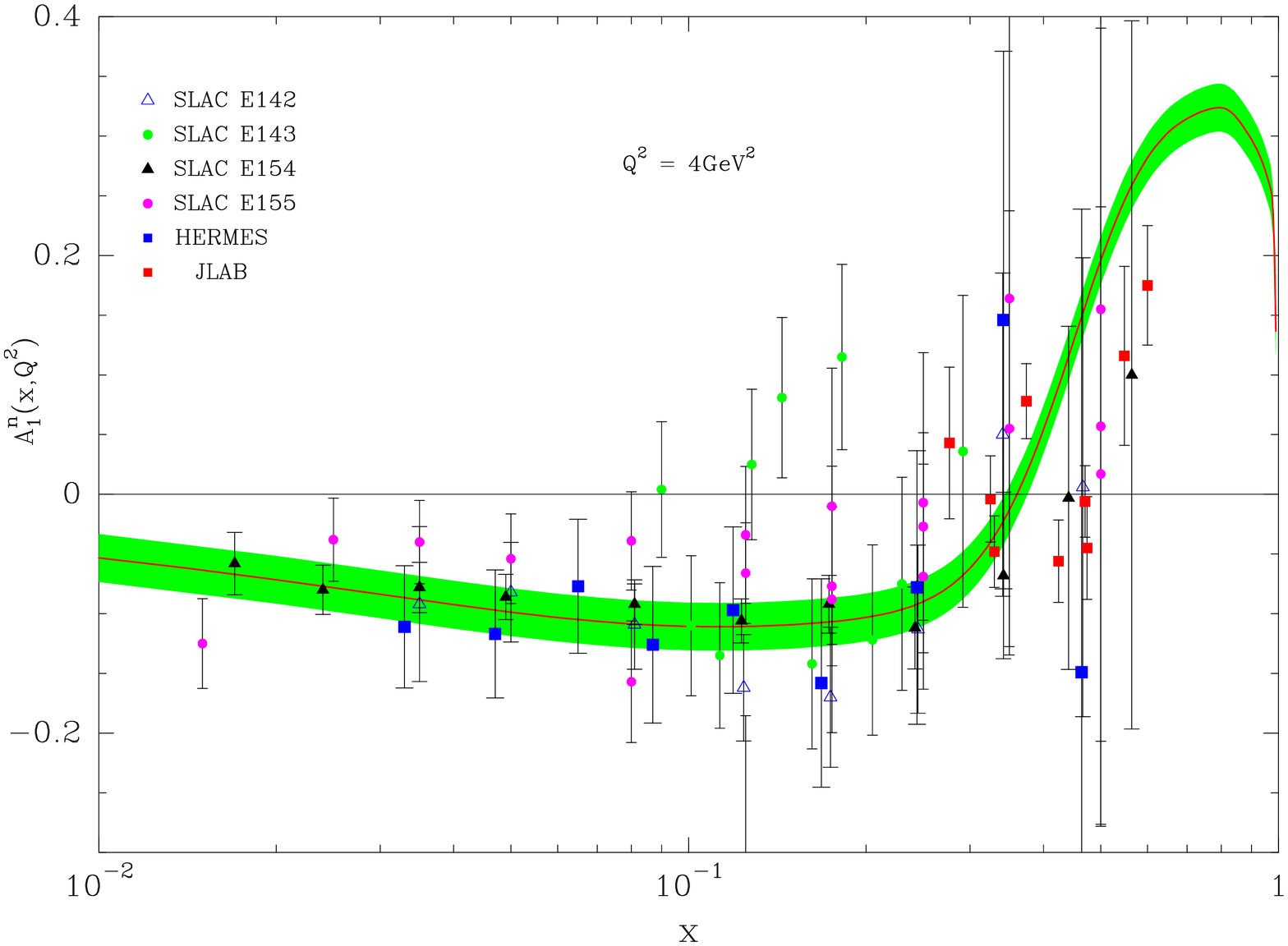}
  \caption[*]{\baselineskip 1pt
 {\it Top}~: Comparison of the world data on $A_1^p(x,Q^2)$ with the result of
the statistical approach at $Q^2= 4~\mbox{GeV}^2$, including the corresponding
error band.\\
{\it Bottom}~: Comparison of the world data on $A_1^n(x,Q^2)$ with the result
of the statistical approach at $Q^2= 4~\mbox{GeV}^2$, including the
corresponding error band. Data are from \cite{herm05} - \cite{clas06}.}
\label{figa1}
\end{center}
\end{figure}

We now turn to the important issue concerning
the asymmetries $A_1^{p,d,n}(x,Q^2)$, measured in polarized DIS.
We recall the definition of the asymmetry $A_1(x,Q^2)$, namely
\begin{equation}
A_1(x,Q^2)= \frac{[g_1(x,Q^2) - \gamma^2 (x,Q^2) g_2(x,Q^2)]}{F_2(x,Q^2)}
\frac{2x[1+R(x,Q^2)]}{[1+\gamma^2(x,Q^2)]}~,
\label{26}
\end{equation}
where $g_{1,2}(x,Q^2)$ are the polarized structure functions,
$\gamma^2(x,Q^2)=4x^2 M_{p}^2/Q^2$ and $R(x,Q^2)$ is the ratio between the
longitudinal and transverse photoabsorption cross sections. When $x \to 1$ for
$Q^2= 4~\mbox{GeV}^2$, $R$ is the order of 0.30 or less and $\gamma^2(x,Q^2)$
is close to 1, so if the $u$ quark dominates, we have $A_1 \sim 0.6 \Delta
u(x)/u(x)$, so it is unlikely to find $A_1 \to 1$, as required by the counting
rules prescription, which we don't impose.
We display in Fig.~\ref{figa1} the world data on $A_1^{p}(x,Q^2)$ ({\it Top})
and $A_1^{n}(x,Q^2)$ ({\it Bottom}), with the results of the statistical
approach at $Q^2= 4 \mbox{GeV}^2$, up to $x=1$. Indeed we find that $A_1^{p,n}
< 1$.

Finally one important outcome of this new analysis of the polarized DIS data
in
the framework of the statistical approach, is the confirmation of a large
positive
gluon helicity distribution, which gives a significant contribution to the
proton spin \cite{bs14}.

\newpage

\section{Hadronic collisions}
A precise determination of parton distributions allows us to use them
as input information to predict strong interaction processes, for 
additional tests of pertubative QCD and also for the search of
new physics. Here we shall test our statistical parton distributions for the  
description of two inclusive reactions, single-jet and $W^{\pm}$ productions
in
$pp$ and $\bar p p$ collisions.

\subsection{Single-jet production in $pp$ and $\bar p  p$ collisions }

\begin{figure}[htbp]    
\vspace*{-15.5ex}
\begin{center}
\includegraphics[width=8.5cm]{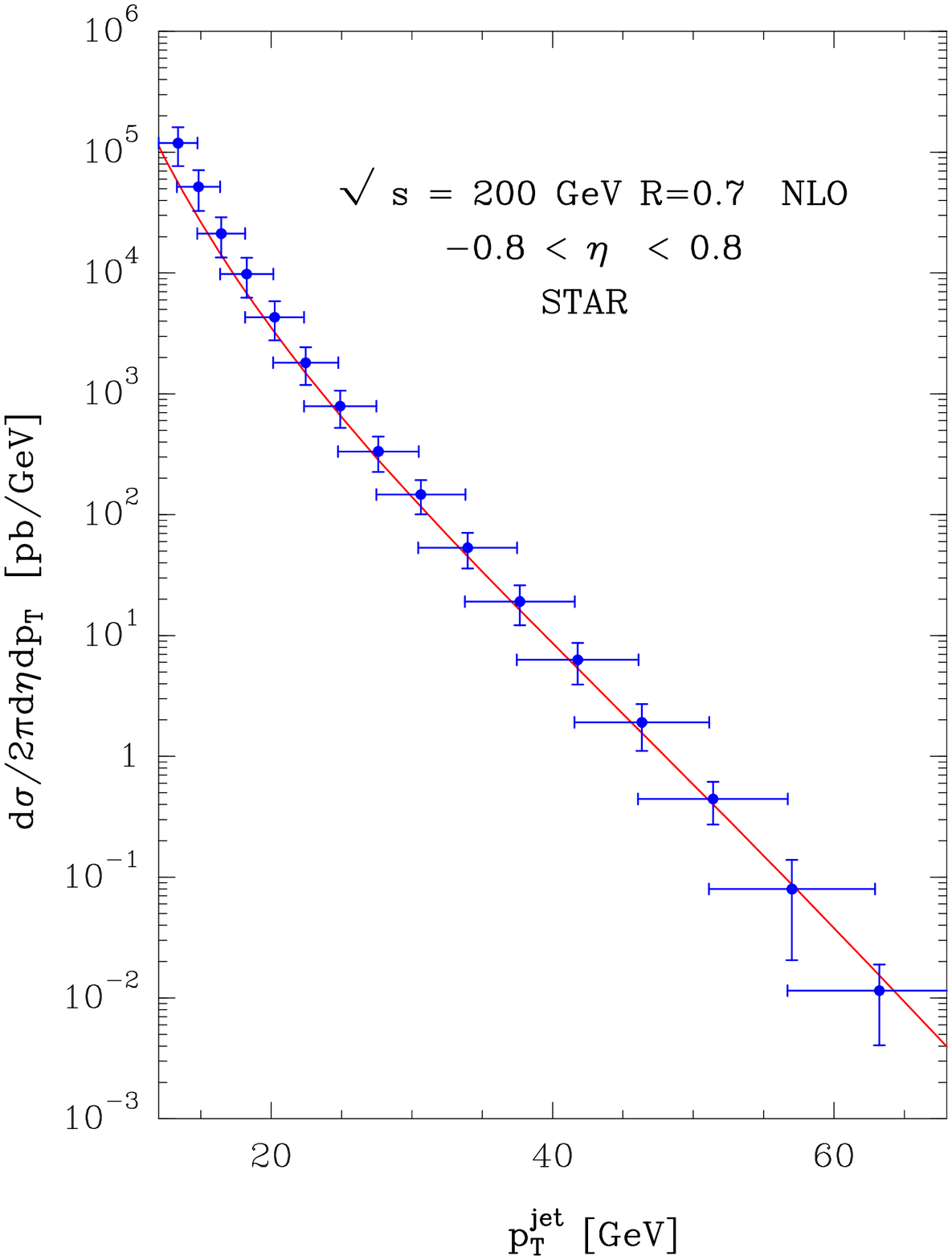}
\vspace*{7.5ex}
\includegraphics[width=10.0cm]{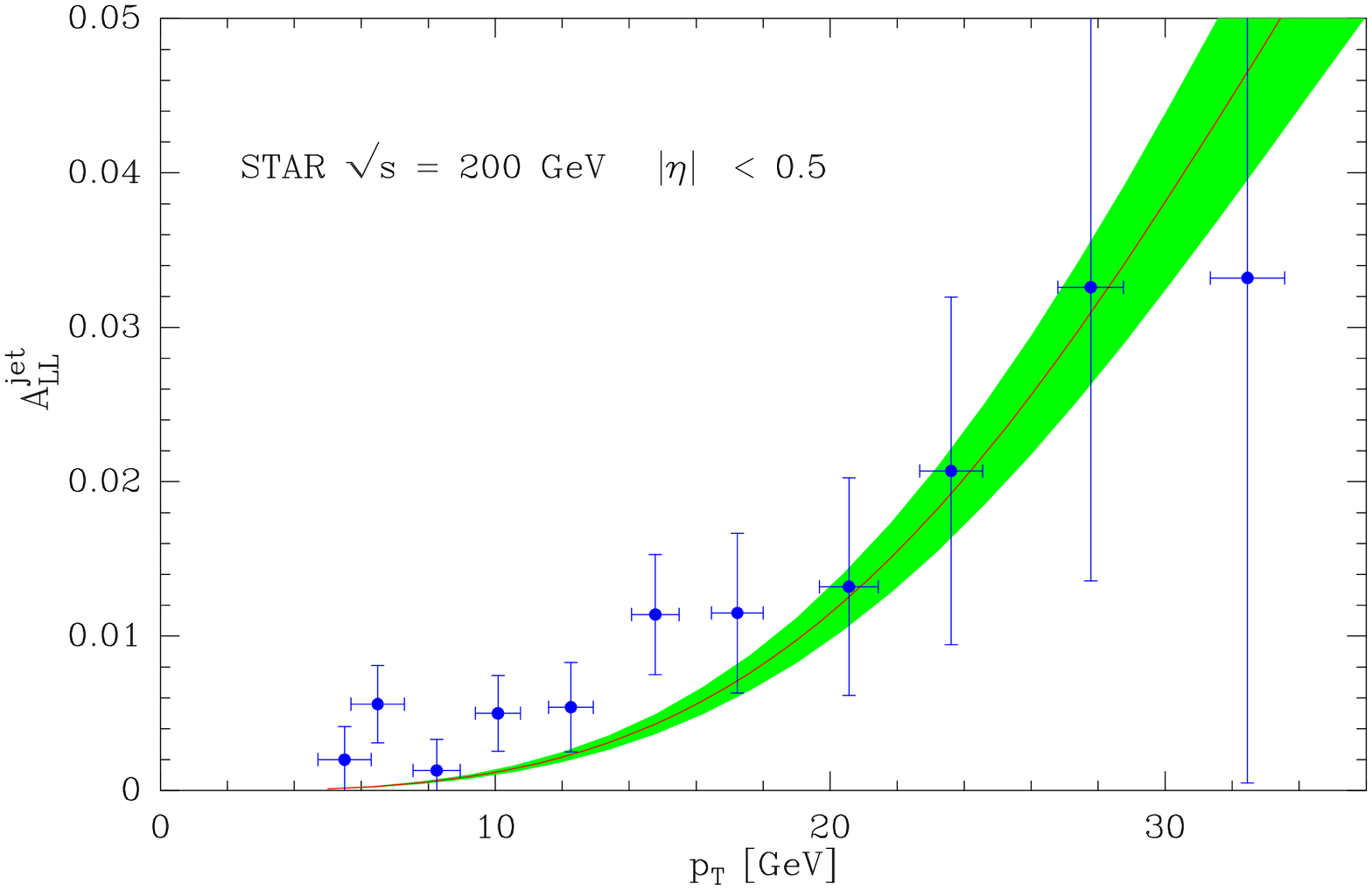}
\vspace*{-3.0ex}
\caption[*]{\baselineskip 1pt
 {\it Top}: Double-differential inclusive single-jet cross section in $ p p$
collisions at $\sqrt{s}$ = 200GeV, versus $p_T^{jet}$, with jet radius
parameter R=0.7, for $-0.8 <  \eta  < 0.8$, from STAR data, obtained with an
integrated luminosity of 5.39pb$^{-1}$  \cite{star1} and the prediction from
the statistical approach.\\
{\it Bottom}:  Our predicted double-helicity asymmetry $A_{LL}^{jet}$ for
single-jet production at BNL-RHIC in the near-forward rapidity region, versus
$p_T$ and the data points from STAR \cite{star}, with the corresponding error
band.}
\label{starjet}
\end{center}
\end{figure}

The cross section for the production of a single-jet of rapidity $y$ and 
transverse momentum $p_T$, in a $pp$ or $\bar{p}p$ collision is given, at
lowest-order (LO), by  
\begin{eqnarray}
E\frac{d^3\sigma}{dp^3} 
&=& \sum_{ij}\frac{1}{1+\delta_{ij}}\frac{2}{\pi}
\int_{x_0}^{1} dx_a \frac{x_a x_b}{2x_a - x_Te^y} \times
\nonumber\\
&&\left[f_i(x_a, Q^2) f_j(x_b, Q^2)\frac{d\hat \sigma_{ij}}{d\hat{t}}(\hat s,
\hat t, \hat u) + (i\leftrightarrow j) \right]~,
\label{crossjet}
\end{eqnarray}
where $x_T = 2p_T / \sqrt{s}$, $x_0 = x_T e^y/(2 - x_T e^{-y})$,
$x_b = x_a x_T e^{-y}/(2x_a - x_T e^y)$ and $\sqrt{s}$
is the center of mass energy of the collision. In the above sum, $i,j$ stand 
for initial gluon-gluon, quark-gluon and quark-quark scatterings, 
$d\hat \sigma_{ij}/d\hat{t}$ are the corresponding partonic cross sections 
and $Q^2$ is the scaling variable.
The NLO QCD calculations were done using a code
described in Ref.~\cite{JSV}, based on a semi-analytical method within the
"small-cone approximation, improved recently with a jet anti-$k_T$ algorithm for 
a better
definition \cite{MV} \footnote{ We thank Werner Vogelsang for providing 
us with the code to make this calculation.}. In Fig.~\ref{starjet}({\it Top})
our results are compared with the data from STAR
experiment at BNL-RHIC and this prediction agrees very well with the 
data.\\
Now we would like to test, in a pure hadronic collision, our new positive
gluon
helicity distribution, mentioned in Section 2. In a recent paper, the STAR
experiment at BNL-RHIC has reported the observation, in single-jet inclusive
production, of a non-vanishing positive double-helicity asymmetry
$A_{LL}^{jet}$ for $5 \leq p_T \leq 30$GeV, in the near-forward rapidity
region
\cite{star}. We show in Fig.~\ref{starjet}({\it Bottom}) our prediction
compared with these high-statistics data points and the agreement is very
reasonable.

\clearpage

\begin{figure}[hbp]  
\vspace*{-15.5ex}
\begin{center}
\includegraphics[width=7.5cm]{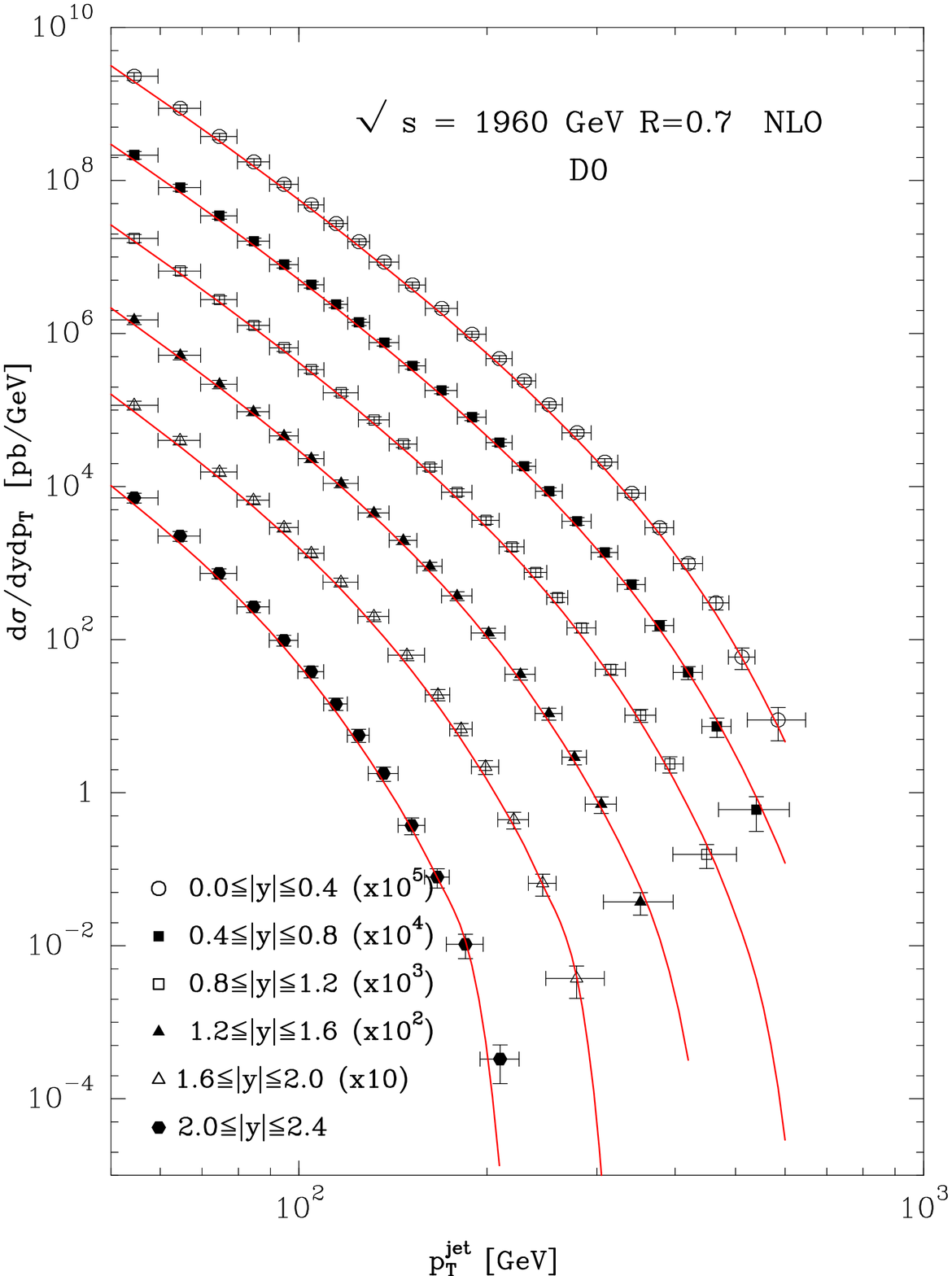}
\includegraphics[width=7.5cm]{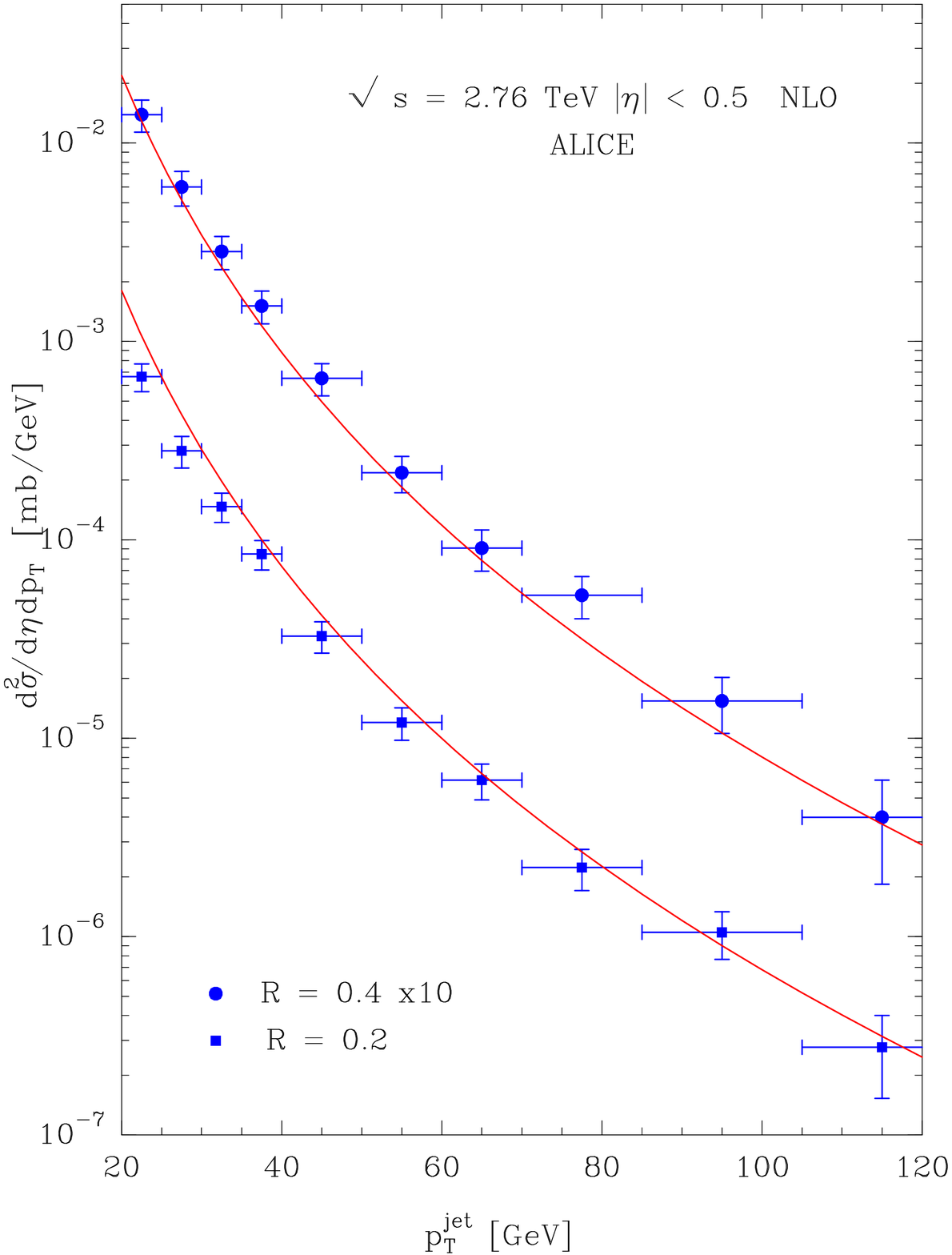}
\caption[*]{\baselineskip 1pt
 {\it Top}: Double-differential inclusive single-jet cross section in $\bar p
p$ collisions at $\sqrt{s}$ = 1.96TeV, versus $p_T^{jet}$, with jet radius
parameter R = 0.7, for different rapidity bins from D0 \cite{d01} and the
predictions from the statistical approach
{\it Bottom}: Same from ALICE \cite{alice} in $pp$ collisions at $\sqrt{s}$ =
2.76TeV, with R = 0.2, 0.4 and $|\eta|< 0.5$.} 
\label{jetd0alice}
\end{center}
\end{figure}

\begin{figure}[hbp]   
\vspace*{-21.5ex}
\begin{center}
\includegraphics[width=7.5cm]{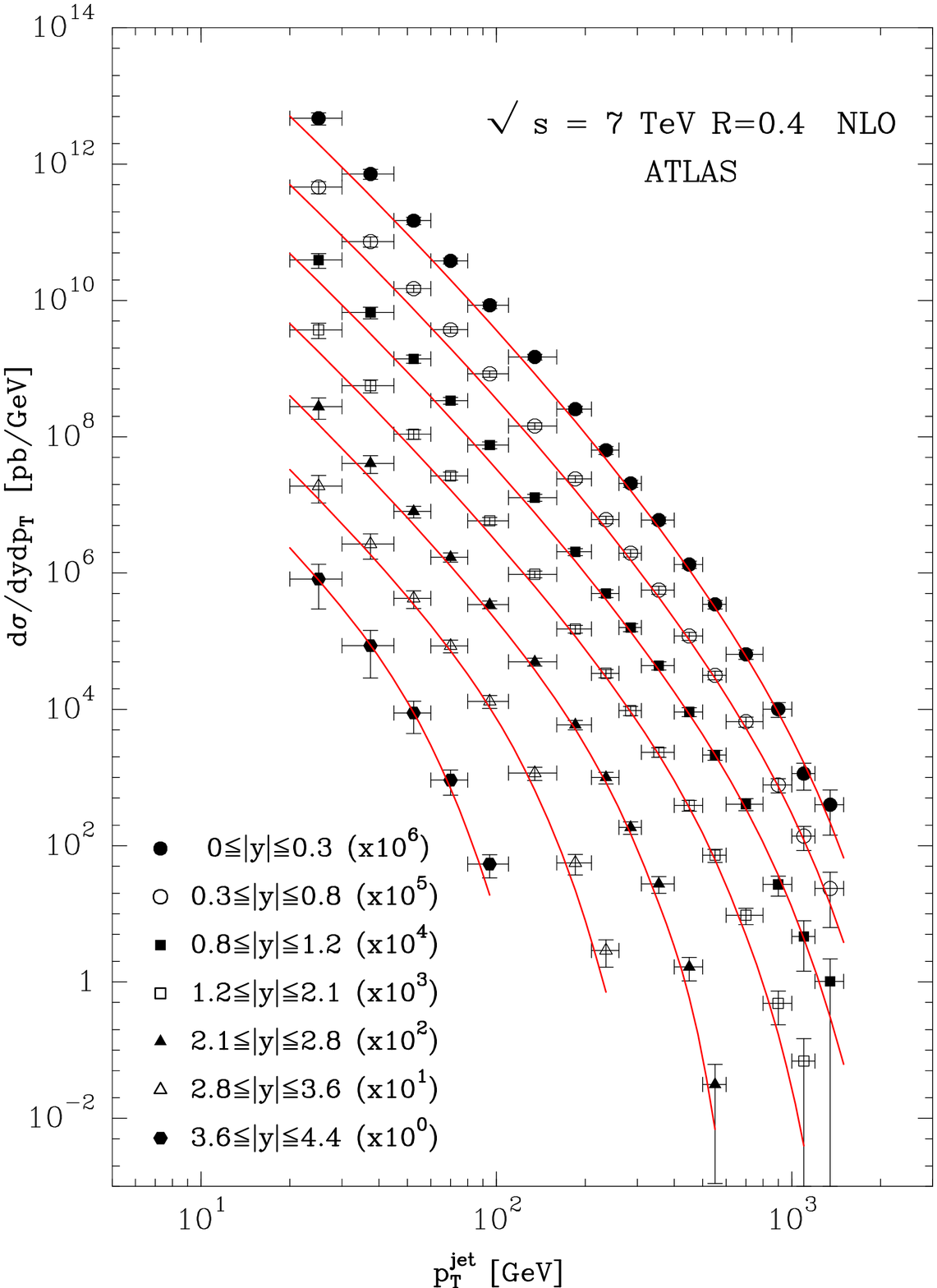}
\includegraphics[width=7.5cm]{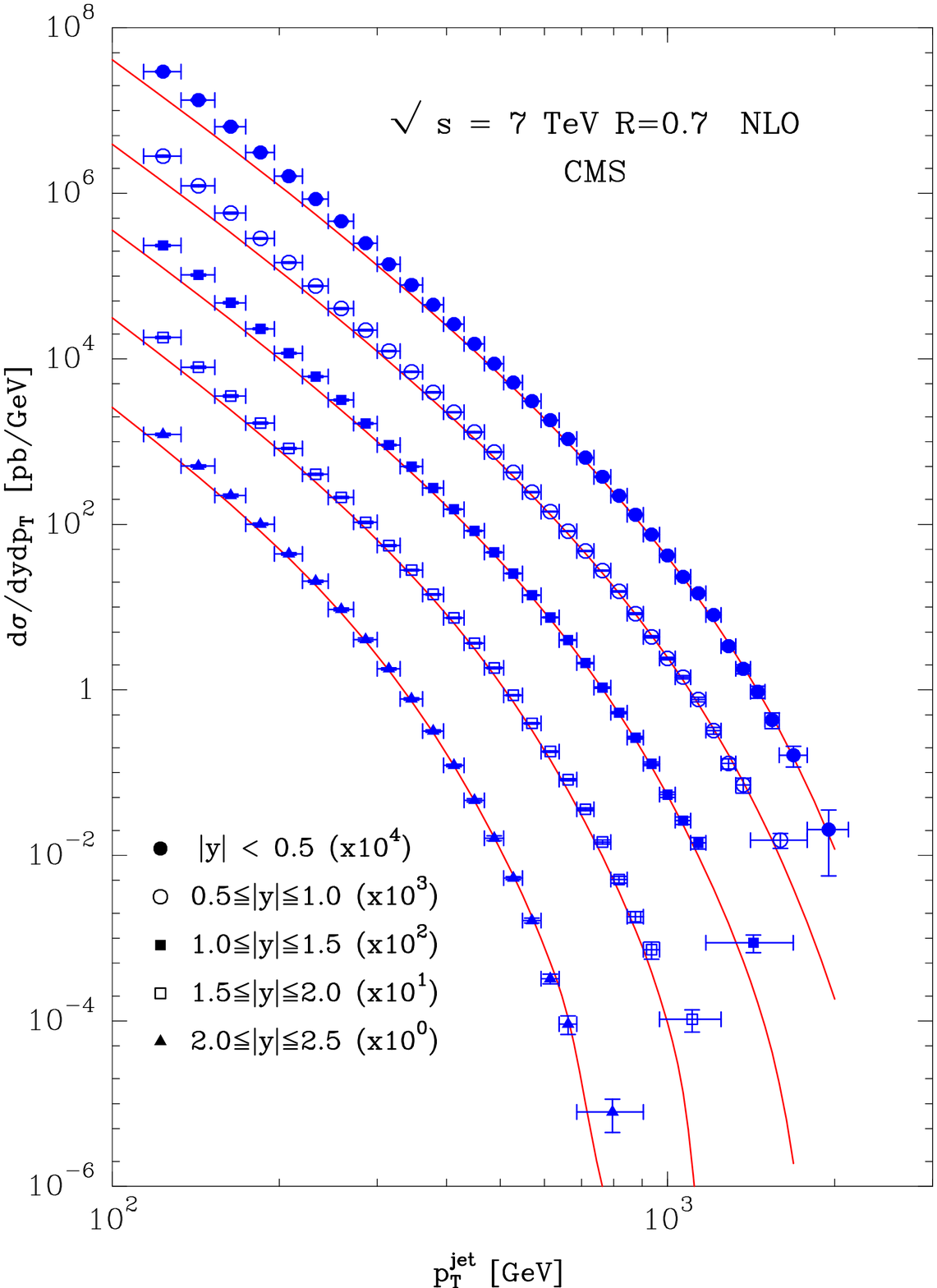}
\caption[*]{\baselineskip 1pt
{\it Top}: Double-differential inclusive single-jet cross section in $pp$
collisions at $\sqrt{s}$ = 7TeV, versus $p_T^{jet}$, with jet radius parameter
R = 0.4, for different rapidity bins from ATLAS \cite{atlasjet} and the
predictions from the statistical approach\\
{\it Bottom}: Same from CMS \cite{cms13}, with R = 0.7.}
\label{jetlhc7}
\end{center}
\end{figure}

\begin{figure}[htbp]  
\vspace*{-21.5ex}
\begin{center}
\includegraphics[width=8.0cm]{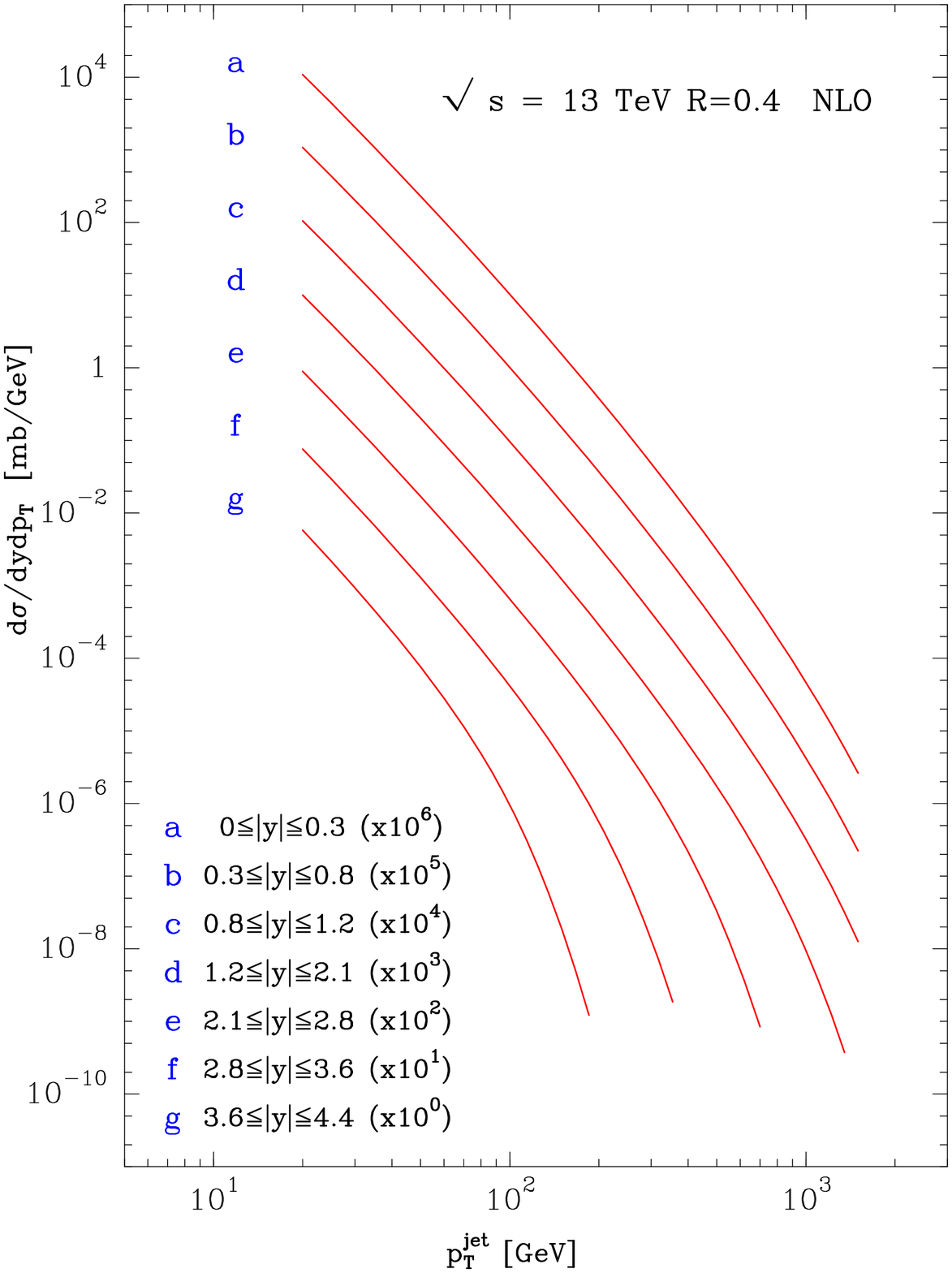}
\caption[*]{\baselineskip 1pt
 Predicted cross sections for single-jet production in $pp$ collisions at
$\sqrt{s}$ = 13TeV, versus $p_T^{jet}$, with jet radius parameter R = 0.4, for
different rapidity bins. }
\label{jetlhc13}
\end{center}
\end{figure}

There are several data sets for the cross section of single-jet production
which will allow us to test our predictions. First we show in
Fig.~\ref{jetd0alice}({\it Top}), the results, versus $p_T^{jet}$ for
different
rapidity bins, from D0 \cite{d01} and the results from ALICE \cite{alice} in
Fig.~\ref{jetd0alice}({\it Bottom}).  Except ALICE, STAR and D0 are in good agreement with the
statistical approach, as well as the results from ATLAS and CMS displayed in
Fig. \ref{jetlhc7} at $\sqrt{s}$ = 7TeV. We are giving in Table 3 the detailed
number of data points and $\chi^2$ for all these experiments. Given the fact
that the experimental
results are falling off over more than 
eight orders of magnitude, this is a remarkable confirmation of the Standard
Model expectations, leaving no room for new physics. However some deviations
might occur when LHC will reach a higher energy regime and this is why we give
in  Fig. \ref{jetlhc13}, our predictions for the single-jet cross section at
$\sqrt{s}$ = 13TeV.

\begin{table}[hbp]
\begin{center}
\begin{tabular}{ c c c c c}
\hline
Experiment  &$\sqrt{s}$ TeV & $\chi^2$ &$N_{data}$ &$\chi^2$/d.o.f. \\
\hline\raisebox{0pt}[12pt][6pt]
STAR \cite{star1} &  0.2 &  11 &  16 &   0.67  \\[4pt]
D0 \cite{d01}  &    1.96& 97 &  110 &  0.88  \\[4pt]
ALICE \cite{alice} & 2.76 & 67 &  20 &   3.38  \\[4pt]
ATLAS \cite{atlasjet} & 7.0 & 127 &  88  &  1.45   \\[4pt]
CMS \cite{cms13} & 7.0 & 374 &  132 & 2.84  \\[4pt]
\hline
\end{tabular}
\caption {Detailed $\chi^2$ prediction for the inclusive single jet
production.}
\label{table3}
\end{center}
\end{table}

\subsection{$W^{\pm}$ production in $pp$ and $\bar p  p$ collisions}
\begin{figure}[hbp]   
\begin{center}
\includegraphics[width=7.5cm]{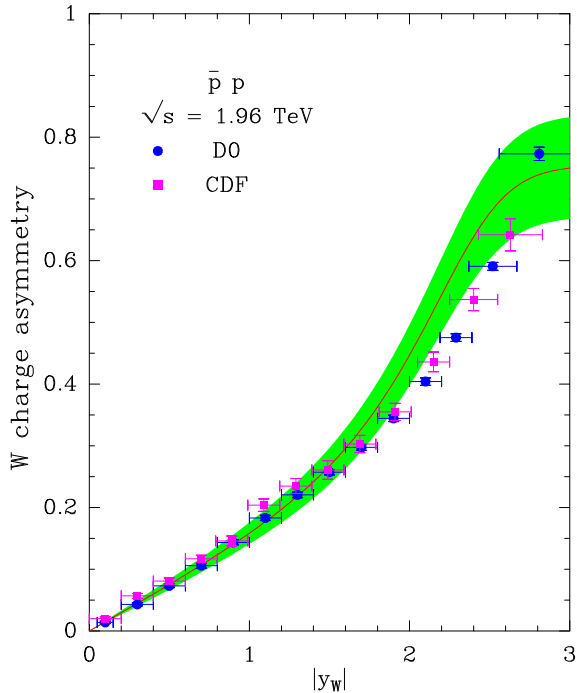}
\caption[*]{\baselineskip 1pt
 The measured $W$ production charge asymmetry at FNAL-Tevatron \cite{cdf,d0},
versus the $W$ rapidity $y_W$ and the prediction from the statistical
approach,
including the corresponding error band.}
\label{asymw}
\end{center}
\end{figure}

\begin{figure}[htp]  
\vspace*{-10.5ex}
\begin{center}
\includegraphics[width=7.5cm]{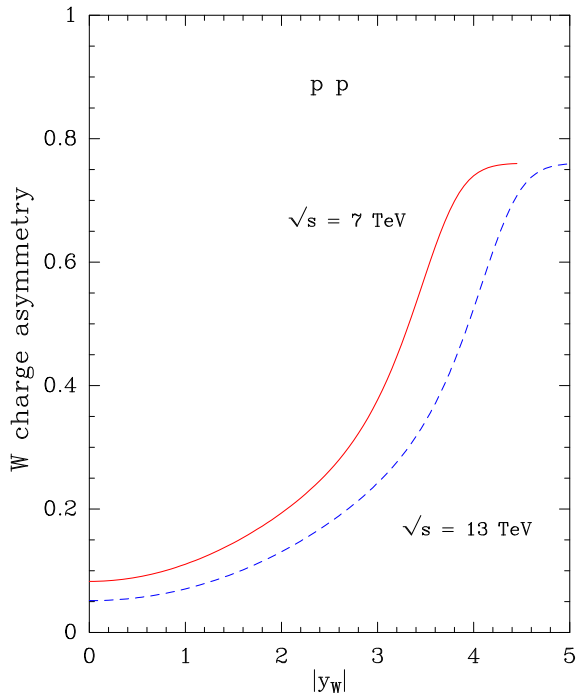}
\caption[*]{\baselineskip 1pt
 The predictions from the statistical approach for the $W$ production charge
asymmetry at the LHC energies, $7 \mbox{TeV}$ (solid line) and $13 \mbox{TeV}$
(dashed line), versus the $W$ rapidity $y_W$. }
\label{asymwlhc}
\end{center}
\end{figure}

 Let us recall that for the $W^{\pm}$ production in $pp$ collision, the
differential cross section $d\sigma_{pp}^{W^{\pm}}/dy$ can be computed
directly
from the Drell-Yan picture in terms of the {\it dominant} quark-antiquark
fusion reactions $u\bar d\to W^+$ and $\bar u d \to W^-$. So for $W^+$
production, we have to LO
\begin{equation}
d\sigma_{pp}^{W^+}/dy \sim u(x_1, M_{W}^2)\bar {d}(x_2, M_{W}^2) + \bar
{d}(x_1, M_{W}^2)u(x_2, M_{W}^2)~,
\label{ppW+}
\end{equation}
where $x_{1,2} = M_W/\sqrt{s} \exp(\pm y)$, $y$ is the rapidity of the W and
$\sqrt{s}$ denotes the $c.m.$ energy of the collision. For $W^-$ production,
we
have a similar expression, after quark flavors interchanged and clearly these
$y$ distributions are symmetric under $y \to -y$. In the case of $\bar p p$
collision we have
\begin{equation}
d\sigma_{\bar p p}^{W^+}/dy \sim u(x_1, M_{W}^2) d(x_2, M_{W}^2) + \bar
{d}(x_1, M_{W}^2)\bar {u}(x_2, M_{W}^2)~,
\label{barppW+}
\end{equation}
which is no longer symmetric under $y\to -y$. However $W^+$ and
$W^-$ production cross sections are simply related since we have $\frac{d\sigma_{\bar p p}^{W^-}}{dy} (y) =
\frac{d\sigma_{\bar p p}^{W^+}}{dy} (-y)$.\\
Let us now turn to the charge asymmetry defined as
\begin{equation}
A(y) = \frac{\frac{d\sigma_{\bar p p}^{W^+}}{dy} (y) - \frac{d\sigma_{\bar p
p}^{W^-}}{dy} (y)}{\frac{d\sigma_{\bar p p}^{W^+}}{dy} (y) +
\frac{d\sigma_{\bar p p}^{W^-}}{dy} (y)}~,
\label{Ay}
\end{equation}
and clearly we have $A(y) = -A(-y)$.\\
It contains very valuable informations on the light quarks distributions
inside
the proton and in particular on the ratio of down-to-up-quark, as noticed long
time ago \cite{bhkw}. Although the cross sections are largely modified by NLO
and NNLO QCD corrections, it turns out that these effects do not affect the LO
calculation of the charge asymmetry \cite{admp}. A direct measurement of this
asymmetry has been achieved at FNAL-Tevatron by CDF  \cite{cdf} and D0
\cite{d0} and the results are shown in Fig. \ref{asymw},  together with the
prediction of the statistical approach. The agreement is very good in the
low-$y$ region. However in the high-$y$ region the charge asymmetry might not
flatten out, following the behavior of our predicted $d(x)/u(x)$ ratio in the
high-$x$ region (see Fig. 4 of Ref. \cite{bbs4}).\\
In view of forthcoming data from the LHC, we display in Fig. \ref{asymwlhc},
predictions from the statistical approach at $\sqrt{s}$ = 7 and 13 TeV. In
this
case the $W$ charge asymmetry is symmetric in $y_W$ and at
fixed $y_W$ it decreases for increasing energy.\\

However,  it is not always possible to reconstruct the $W$-boson and to
measure
the boson rapidity, because
of the energy carried away by the neutrinos in leptonic $W$-boson decays.
A quantity more directly accessible experimentally
is  the lepton charge asymmetry, defined as
\begin{equation}
 A(\eta) = \frac{d\sigma/d\eta(W^+\to l^{+} \nu)-d\sigma/d\eta(W^-\to l^- \bar
\nu)}{d\sigma/d\eta(W^+\to l^{+} \nu) + d\sigma/d\eta(W^-\to l^- \bar \nu)},
\end{equation}
where  $d\sigma/d\eta$ is the differential
cross section for $W$-boson production and subsequent leptonic decay and
$\eta= -\ln{[\tan{(\theta/2)}]}$ is the charged lepton
pseudorapidity in the laboratory frame, with
 $\theta$ being the polar angle measured with respect to the beam axis.

\begin{figure}[htp]   
\begin{center}
\includegraphics[width=11.0cm]{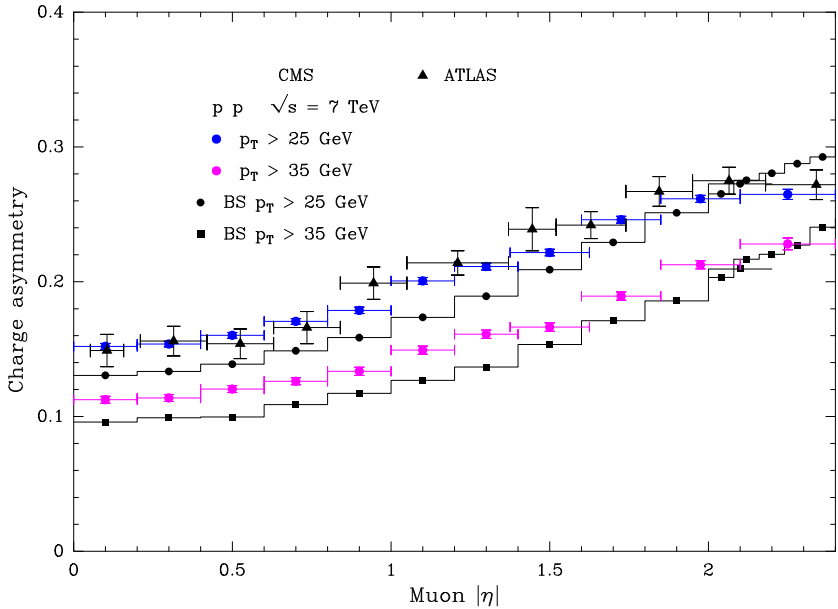}
\caption[*]{\baselineskip 1pt
 The  $\mu$ charge asymmetry from $W$-boson decays  in bins of absolute muon
pseudorapidity at the LHC  $7 \mbox{TeV}$, with some kinematical cuts
$p_T^{\mu} > $  20GeV for ATLAS \cite{atlasw} and $p_T^{\mu} > $   25, 35GeV
for CMS \cite{cms14} with the predictions of the statistical approach (BS). }
\label{asymmulhc}
\end{center}
\end{figure}

There was an earlier experimental result at the LHC from ATLAS \cite{atlasw},
obtained with a total integrated luminosity of 31pb$^{-1}$, but more recently
CMS \cite{cms14} released a data sample corresponding to
a total integrated luminosity of. 4.7fb$^{-1}$. We display in
Fig.~\ref{asymmulhc} both data sets, together with the results of our
calculations, which were obtained  using the FEWZ code \cite{fewz}. Although
the
statistical approach is compatible with CMS data, a higher accuracy is
required
before considering that it is a very conclusive test of our PDF's.\\

Finally we consider the process $\overrightarrow p p\to W^{\pm} + X \to
e^{\pm}
+
X$, where the arrow denotes a longitudinally polarized proton and the outgoing
$e^{\pm}$ have been produced by the leptonic decay of the $W^{\pm}$-boson. The
parity-violating  asymmetry is defined as
\begin{equation}
A_L^{PV} = \frac{d\sigma_+ - d\sigma_-}{d\sigma_+  + d\sigma_-}~.
\label{AL}
\end{equation}
Here $\sigma_h$ denotes the cross section where the initial proton has
helicity
$h$. It is an excellent tool for pinning down the quark helicity
distributions,
as first noticed in Ref. \cite{bs93}.

\begin{figure}[hbp]   
\begin{center}
\includegraphics[width=11.0cm]{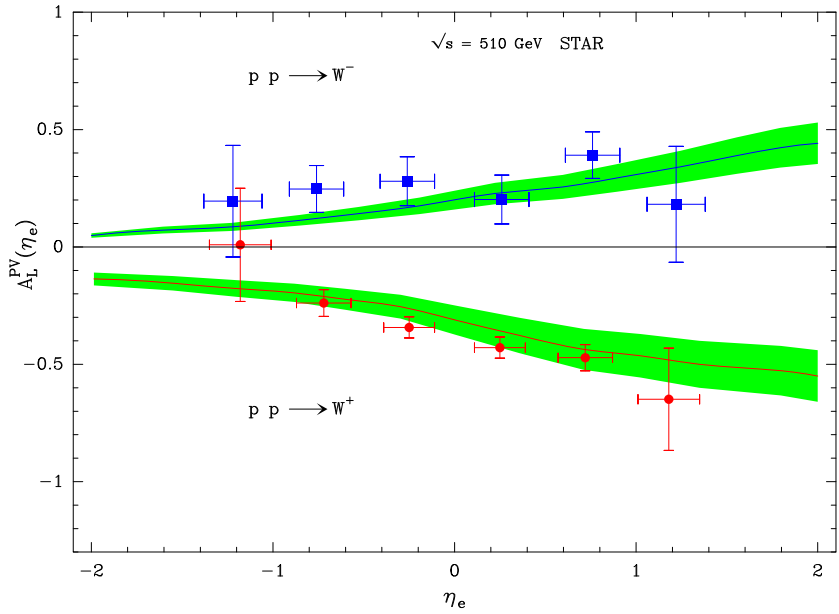}
\caption[*]{\baselineskip 1pt
 The measured parity-violating helicity asymmetries $A_L^{PV}$ for
charged-lepton production at BNL-RHIC from STAR \cite{star14}, through
production and decay of $W^{\pm}$-bosons versus $y_e$  the charged-lepton
rapidity. The solid curves are the predictions of the statistical approach,
including the error bands. }
\label{alpv}
\end{center}
\end{figure}

$A_L^{PV}$ was measured recently at RHIC-BNL \cite{star14} and the results are
shown
in Fig.~\ref{alpv}. As explained in Ref. \cite{bbsW}, the $W^-$
asymmetry is very sensitive to the sign and magnitude of $\Delta \bar u$, so
this is another successful result of the statistical approach.


\clearpage

\section{Concluding remarks}
Our quantum statistical approach to parton distributions, proposed thirteen
years ago, has been revisited in the light of 
a large set of most recent world data from spin-averaged and spin-dependent
pure DIS experiments, excluding semi-inclusive DIS and hard scattering
processes. The construction of the PDF allows us
to obtain simultaneously the unpolarized distributions and the helicity
distributions, a rather unique situation. In the current literature one finds,
on the one hand, global QCD analysis of spin-averaged experiments, some
including LHC data, \cite{h112,mrst,nnpdf,ct10,abm,jdr} to extract unpolarized
distributions and, on the other hand, global QCD analysis of spin-dependent
experiments \cite{dssv1,others,dssv2,nocera}, to determine the helicity
distributions. They don't restrict themselves to DIS experiments, like we do,
and in general their parameterizations involve many free parameters, whose
total number and physical meaning are considered to be irrelevant.

Our aim in the analysis of the DIS data was to incorporate in the structure of
the PDF some physical principles, like Bose-Einstein distributions for gluons
and Fermi-Dirac distributions for quarks and antiquarks, which are simply
related from chiral properties of QCD. This allows us to reduce the number of
free parameters, some of them having a physical interpretation. The
improvements we have obtained from this new version is a more accurate
determination of light quarks, strange quarks, strongly related to their
corresponding antiquarks, and gluon distributions. We have found a large
positive gluon helicity distribution, $\Delta G(x,Q^2)$, very similar to that
coming from the results of Ref.~\cite{dssv2} (see also Ref.~\cite{nocera}) and
compatible with a new high energy hadron collider result. As
we have seen there are several challenging questions related to large-$x$
predictions, because the large-$x$ structure of hadrons is essential for a
complete picture.

The predictive power of our approach lies partly in the DIS sector, but mainly
in the rich domain of hadronic collisions, up to LHC energies. We have shown
that our predictions for several spin-averaged and spin-dependent processes at
RHIC, Tevatron and LHC, are already in fair agreement with existing data and
we
expect this will be confirmed by forthcoming experiments.

\end{document}